\documentclass[aps,pre,showpacs,floatfix,twocolumn,superscriptaddress]{revtex4-2}

\usepackage{mathrsfs}
\usepackage[ 
 colorlinks=true,
 pdfborder={0 0 0},
 linkcolor=red,
 citecolor = red
]{hyperref}
\usepackage{graphicx}
\usepackage{bm,amsmath,amssymb,accents}
\newcommand{\myvect}[1]{\accentset{\rightharpoonup}{#1}}
\usepackage{dcolumn}
\usepackage{xcolor}
\usepackage{bm}

\begin{document}

\title{Machine learning for percolation utilizing auxiliary Ising variables}

\author{Junyin Zhang}
\thanks{These two authors contributed equally to this paper.}
\affiliation{
Hefei National Laboratory for Physical Sciences at the Microscale, University of Science and Technology of China, Hefei 230026, China}
\affiliation{Department of Modern Physics, University of Science and Technology of China, Hefei 230026, China}

\author{Bo Zhang}
\thanks{These two authors contributed equally to this paper.}
\affiliation{
Hefei National Laboratory for Physical Sciences at the Microscale, University of Science and Technology of China, Hefei 230026, China}
\affiliation{Department of Modern Physics, University of Science and Technology of China, Hefei 230026, China}

\author{Junyi Xu}

\affiliation{College of Physics and Optoelectronics, Taiyuan University of Technology, Shanxi 030024, China} 
\author{Wanzhou Zhang}
\thanks{Corresponding author: zhangwanzhou@tyut.edu.cn}
\affiliation{CAS Key Laboratory of Quantum Information, University of Science and Technology of China, Hefei 230026, China}
\affiliation{College of Physics and Optoelectronics, Taiyuan University of Technology, Shanxi 030024, China} 
\affiliation{
Hefei National Laboratory for Physical Sciences at the Microscale, University of Science and Technology of China, Hefei 230026, China}
\author{Youjin Deng}
\thanks{Corresponding author: yjdeng@ustc.edu.cn}
\affiliation{
Hefei National Laboratory for Physical Sciences at the Microscale, University of Science and Technology of China, Hefei 230026, China}
\affiliation{Department of Modern Physics, University of Science and Technology of China, Hefei 230026, China}

\affiliation{Shanghai Research Center for Quantum Sciences, Shanghai 201315, China}

\affiliation{MinJiang Collaborative Center for Theoretical Physics, College of Physics and Electronic Information Engineering, Minjiang University, Fuzhou 350108, China}

\date{\today}

\begin{abstract}
Machine learning for phase transition has received intensive research interest in recent years. However, its application in percolation still remains challenging. We propose an auxiliary Ising mapping method for the machine learning study of the standard percolation as well as a variety of statistical mechanical systems in correlated percolation representation. We demonstrate that unsupervised machine learning is able to accurately locate the percolation threshold, independent of the spatial dimension of system or the type of phase transition, which can be first-order or continuous. Moreover, we show that, by neural network machine learning, auxiliary Ising configurations for different universalities can be classified with a high confidence level. Our results indicate that the auxiliary Ising mapping method, despite its simplicity, can advance the application of machine learning in statistical and condensed-matter physics.

\end{abstract}

\maketitle
\section{introduction}
\label{section:introduction}
 \begin{figure*}
 \centering
 \includegraphics[width=0.95\textwidth]{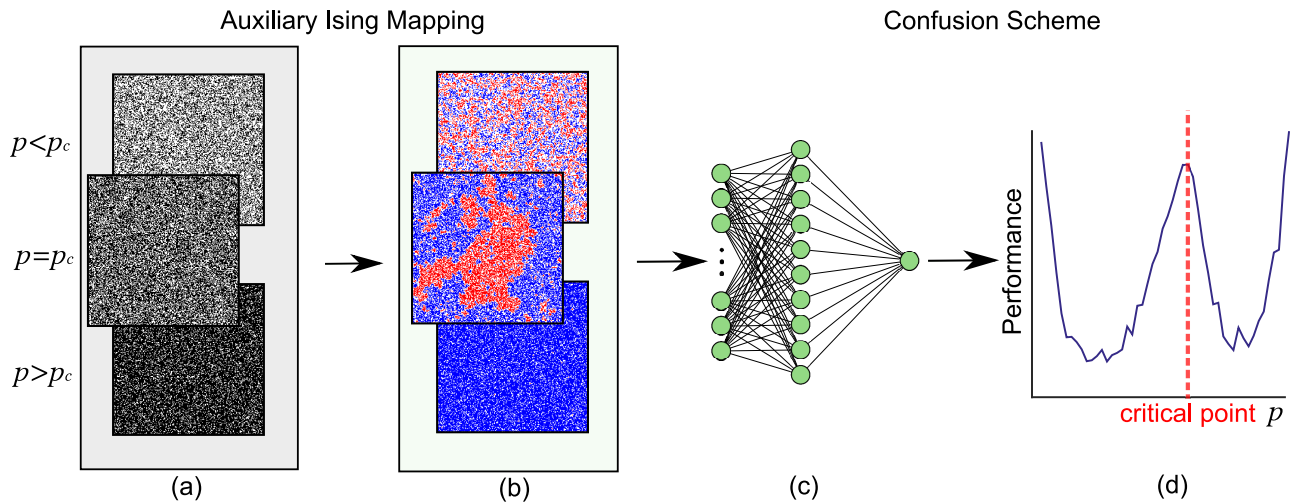}
 \caption{The main idea of this work. Auxiliary Ising mapping (AIM) is introduced to transform the original percolation configuration into an Ising-like spin configuration, while all the correlation functions are kept unchanged. After applying the AIM, the percolation phase transition can be identified in an unsupervised way using a confusion scheme based on the neural network, the details of the network are in the Appendix~\ref{subsection:appendix--Binary_Classification_for_confusion}. (a) The three snapshots from top to bottom are the original configurations (in black and white colors) of the site percolation on the square lattice with occupation probabilities $p=0.4$, $0.593$ (the percolation threshold) and $0.8$, respectively. (b) The corresponding configurations obtained after AIM. The auxiliary Ising spins are colored in blue and red, and the unoccupied sites are colored in white. The percolation clusters emerged as the $p$ increases. (c) The fully connected neural network, including one input layer, one hidden layer, and one output neuron, works as a binary classification neural network. (d) The W-shaped confusion performance curve with the central peak locates at the percolation threshold indicates the percolation transition.}
 \label{fig:first_diagram}
\end{figure*}

The percolation model was first proposed as a model for a porous medium by Broadbent and Hammersley in 1957~\cite{bm} and then has been applied in physics,
materials science, epidemiology, finance and other fields~\cite{5,6,duffie2007information,1986Percolation}.
~As a paradigm of random and semi-random connectivity, percolation models on regular lattices or complex network have played a key role in statistical mechanics and network science~\cite{8,9,0Explosive,2021Percolation}. Standard percolation is defined as the random and independent occupation of each site or bond on a lattice or complex network according to a certain probability. Furthermore, correlated percolation has been also proposed\cite{fk,fk_two,fisher1967physics,coniglio1980clusters,hu} to capture the thermal properties of various physical models, in which the occupation probability of each site or bond is not independent. The concept of correlated percolation has been used later in the famous Swendsen-Wang algorithm~\cite{PhysRevLett.58.86}.

Recently, machine learning has provided new ideas and methods for studying various physical problems~\cite{review}, among which the phase transition is one of the fruitful topics. Machine learning can generally be divided into supervised and unsupervised learning. Besides the extensive works by supervised phase detection, many studies based on unsupervised phase detection have also been reported recently, such as the principal component analysis (PCA)~\cite{pca1,pca2,pca3}, the t-Distributed Stochastic Neighbor Embedding (t-SNE) ~\cite{ Zhang_W, tsne4, tsne3}, the diffusion map~\cite{df1, df2, df3,df4} and the confusion scheme ~\cite{confusion} \textit{et al}. However, it is still a challenge for unsupervised machine learning to detect the percolation phase transition points properly~\cite{huizhai,Zhang_W,hzhao}. It was found that that~\cite{huizhai}, after reducing the dimensionality of the original configuration using the PCA method, no statistically significant correlations can be found between the principal components and the order parameters. As far as the confusion method, the performance curve appeared to have a V-shape instead of a W-shape~\cite{hzhao}. Regarding the t-SNE method, it also fails to learn the phase transition of the percolation model, although it can clearly separate the configurations far away from the percolation threshold~\cite{Zhang_W}.

The difficulty in applying machine learning methods to percolation model lies in two parts. On the one hand, the lattice (graph) structures play a crucial role in percolation model and must be encoded in the training set, otherwise the critical behaviors of percolation model can not be captured. On the other hand, percolation is a global property which is difficult to be captured. For similar reasons, the unsupervised phase detection on the XY model is only possible when the data is pre-classified by different global winding numbers~\cite{PhysRevResearch.3.013074,df1}.

Motivated by the previous research works~\cite{Zhang_W,hzhao,huizhai}, here we propose the auxiliary Ising mapping (AIM) method, which maps the original percolation configuration onto an Ising-like spin configuration. The AIM method preserves all correlation functions unchanged and captures the network structure. After such a mapping, we apply the confusion scheme to the spin configurations and find that the percolation threshold can be detected in an unsupervised way. The scheme and a typical result are shown in Fig.~\ref{fig:first_diagram}. Beyond uncorrelated percolation model, we successfully extract the thermal phase transitions of the Potts model and obtain a similar result as Refs.~\cite{wessel,cf_qlr} for the XY model in the representation of correlated percolation clusters. Our method is generally applicable to any other model as long as it can be formulated in the framework of percolation.

The present paper is organized as follows:
Section ~\ref{section:percolation} introduces the AIM method and successfully applies it to percolation models on various lattices and complex networks. Then, we promote our method to learn the phase transitions of the generalized percolation models like the Potts model and the XY model. Section~\ref{section:Identification_of_universal_classes} demonstrates that the configurations of the different models at the phase transition point could be classified by the neural network after the AIM transformation. Section~\ref{section:Summary_and_Discussion} gives the summary and future perspective.

\begin{figure}
 \includegraphics[width = 0.30\textwidth]{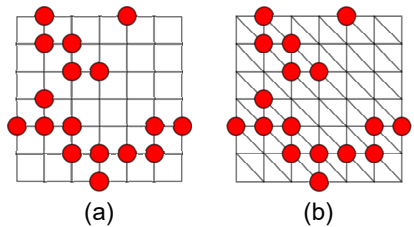}
		\caption{The occupation configurations on (a) the square and (b) the triangular lattice for the site percolation models. Occupied and non-occupied sites are indicated by red symbols and empty sites, respectively. For these two configurations, if only the occupancy configurations are fed in, the neural network can not learn the differences between the different lattices (graphs) and therefore will not learn percolation transition correctly.
		 }
		\label{fig:Tr_Sq_illus}
	\end{figure}

\section{Percolation Transition Detection}
\label{section:percolation}

Identifying phase transitions in a site percolation model through unsupervised machine learning is a challenge~\cite{Zhang_W,hzhao,huizhai}. The problem lies mainly in how to feed the lattice structure of the percolation model as information into the neural networks and other machine learning tools. Suppose we only input the occupation configurations (as in Refs.~\cite{Zhang_W,hzhao,huizhai}) into the neural network, the structure information of the lattice is largely discarded, and the network is unlikely to identify the difference between percolation models in different lattices in an unsupervised way, as illustrated in Fig.~\ref{fig:Tr_Sq_illus}. This difficulty can be solved by the AIM method introduced below, in which the structure information of the lattice and the occupation configurations are combined.

\subsection{Auxiliary Ising Mapping}
\label{subsection:percolatioin--Auxiliary_Ising_Mapping}

As mentioned in Sec.~\ref{section:introduction}, the confusion method can locate the critical point of the thermodynamic phase transition for the Ising model, but does not work for the percolation phase transition. The essential reason is that the status (occupied or unoccupied) of each site in the percolation configuration is generated independently, while the spin status on each site for the Ising model is generated correlatedly. Therefore, a mapping method that can include connectivity information between adjacent sites is helpful.
%according to the Boltzmann distribution followed by the interacting Hamiltonian describing the interaction between spins on different sites. Therefore, it is better to 
%}

%\textcolor{red}{In the mapping, the configuration of clusters is generated based on the occupancy status of each site and the connection information between different sites, and then a dilution-like Ising model configuration is generated, i.e., sites on different clusters are assigned +1 or -1 with equal probability, while sites that do not belong to any cluster are assigned 0. A detail description of the algorithm is as follows}

\begin{figure}[t]
 \centering
 \includegraphics[width=0.30\textwidth]{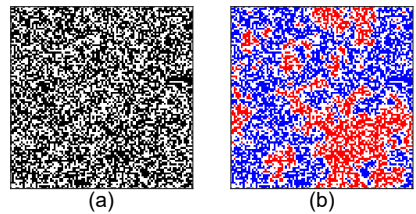}
 \caption{The auxiliary Ising mapping for the square-lattice site percolation. (a) In the initial percolation configuration, occupied sites are marked black, while unoccupied sites are marked white. (b) The AIM configuration.
 These clusters by the occupied sites are randomly labeled $+1$ and $-1$ according to a probability of $1/2$.
 }
 \label{fig:algorithm}
\end{figure}

%\textcolor{blue}{
The algorithm for this mapping is described as follows.
To generate a site percolation configuration with an occupation probability of $p$, the standard procedure is to visit each lattice site sequentially, draw a uniform random number ${\bf rand()} \in [0,1)$, and assign a variable $s=1$ to the site if ${\bf rand()} <p$ and otherwise, $s=0$. The AIM method extends the variable from $s \in \{0,1\}$ to $\{0, \pm1 \}$ and generates site percolation configurations in an epidemic-spreading way. All the sites are initialized to be nonvisited, and, by sequentially visiting the lattice, the first occupied site, called the seed site, is infected by diseases $s =\pm 1$ with equal probability. The seed site then spreads its disease to each of the neighboring and unvisited sites with probability $p$. Note that a visited site, no matter it is empty or already sick, can no longer be infected by the disease. The epidemic keeps spreading until all its boundary sites are visited, and a cluster of infected sites with disease $s=+1$ or $-1$ is formed. Afterward, one sequentially visits the remaining unvisited sites until the next seed site is obtained. The procedure is repeated till all the lattice sites are visited. A configuration of dilute Ising variables $\{ s=0, \pm 1 \}$ is generated, as shown in Fig.~\ref{fig:algorithm}.
%}

%V2: The motivation behind this mapping is the following: it has been shown that the confusion method can be applied to detect the phase transition in Ising-model unsupervised, and the mapping we proposed here, which is inspired by the relation between random cluster model and percolation model, can map the percolation configuration into some Ising like configuration. Besides, as stated in the previous discussion, the problem for applying ML to percolation is the loss of the connection information of the networks. }
With the auxiliary Ising variable, the connectivity information is automatically included in the epidemic spreading process.
%蓝色的逻辑上应该在算法之后，因为提到了epidemic之类的。
%\textcolor{blue}{The algorithm is as follows}:(1) \textbf{Creating configurations:}each site with in order a given occupied probability $p$.(2) \textbf{Forming clusters:}For the occupied sites, the same clusters.The unoccupied sites are not belong to any clusters. (3)\textbf{Introducing auxiliary Ising spins: We randomly assign $1$ or $-1$ to each cluster with equal probability, all sites in the same cluster have the same spin value, and all unoccupied sites are marked by 0. %\textcolor{red}{discuss, shall we add the algorithm?}
Actually, it can be proved that the above algorithm also ensures that the correlation function $G$ remains unchanged between the percolation and the AIM spin configurations. For a given lattice (graph) structure and the occupation probability $p$, an ensemble of percolation configurations $\{\mathcal{C}\}$ can be generated. The correlation function $G(\myvect{r_i},\myvect{r_j})$ refers to a probability that two sites at locations $\myvect{r_i}$ and $\myvect{r_j}$ belong to the same cluster and is defined as the following:
 \begin{equation}
 G(\myvect{r_i},\myvect{r_j}) = \langle \Theta(\mathcal{C};\myvect{r_i},\myvect{r_j})\rangle_{\mathcal{C}}\ ,
 \end{equation}
 where the correlation indicator defined for one specific configuration $\mathcal{C}$ is:
\begin{equation}
 \Theta(\mathcal{C};\myvect{r_i},\myvect{r_j}) = \left\{
 \begin{aligned}
 &0, \qquad \myvect{r_i},\myvect{r_j}\ \text{in\ different\ clusters}\\
 &1, \qquad \myvect{r_i},\myvect{r_j}\ \text{in\ the\ same\ clusters}
 \end{aligned}\right.\ .
\end{equation}
The $\langle\cdots\rangle_\mathcal{C}$ refers to the average of all occupation configurations according to statistical weights. For a percolation configuration with $N_c$ clusters, there are $2^{N_c}$ corresponding auxilliary Ising configurations. A specific auxiliary Ising mapping $AIM_\alpha$ can be denoted by the Ising spins specified on the $N_c$ clusters $\alpha:\{\tilde{s}_1, \cdots, \tilde{s}_{N_C}\}$. 
Different mappings can be labeled as 
\begin{equation}
 AIM_\alpha:\qquad \mathcal{C}\stackrel{\alpha}\longrightarrow\mathcal{S},
\end{equation}
and they lead to various spin configurations $\mathcal{S}$
 on a lattice with $N$ sites, i.e., $\{s_1, \cdots, s_{N}\}$, by which
the correlation between two spins $\langle s_i s_j\rangle_{\alpha}$ still equals to $\Theta(\mathcal{C};\myvect{r_i},\myvect{r_j})$. This can be understood in the following way. If the sites at $\myvect{r_i}$ and $\myvect{r_j}$ belong to the different clusters, then the correlation satisfy the relation 
$\langle s_i s_j\rangle_\alpha $=$\langle s_i\rangle_\alpha \langle s_j\rangle_\alpha =0$ because the different components of $\{\tilde{s}_1, \cdots, \tilde{s}_{N_C}\}$ are independent
from each other. Otherwise the equation
$ s_i s_j =(\pm 1)^2=1$ always holds.

Therefore, the two-point correlation functions of the percolation and spin models are equivalent and can be expressed as follows
\begin{equation}
 G(\myvect{r_i},\myvect{r_j}) = \langle \Theta(\mathcal{C};\myvect{r_i},\myvect{r_j})\rangle_{\mathcal{C}}
 = \langle \langle s_i s_j\rangle_\alpha \rangle_\mathcal{C} = 
 \langle s_i s_j \rangle_{\mathcal{S}}\,
\end{equation}

 \begin{figure*}[htpb]
 \centering
 \includegraphics[width=1.0\textwidth]{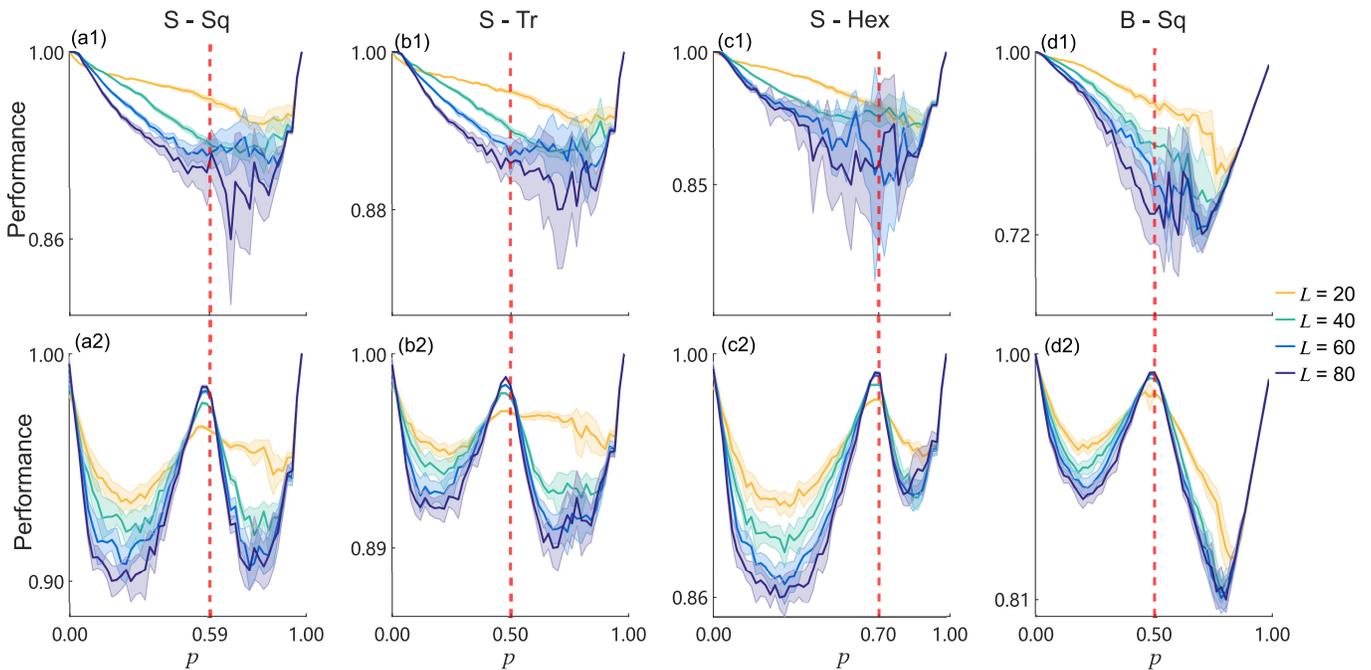}
 \caption{The confusion performance curve for site and bond percolation models. The shaded area denotes the $3\sigma$ errorbar. The curves with (a1-d1) original occupation configurations and (a2-d2) the configurations after AIM on (a) the square-lattice site percolation (S-Sq), (b) the triangular-lattice site percolation (S-Tr), (c) the hexagonal-lattice site percolation (S-Hex) and (d) the square-lattice bond percolation (B-Sq). The red dashed line denotes the percolation threshold. For the square lattice and triangular lattice, the number of sites $N$ is related to linear size $L$ by $N=L^2$, and for the hexagonal lattice it is $N=2L^2$. The periodic boundary condition is applied to the lattices.}
 \label{fig:all_in}
\end{figure*}

, where $\langle \cdots \rangle_\alpha$ and $\langle \cdots \rangle_\mathcal{C}$ refer to averaging over the ensemble of configurations and mappings, respectively. In addition to correlation itself, the AIM algorithm preserves all physical quantities that can be expressed as correlation function and thus the critical behavior of the percolation model. %wz: all ->add restrict
As an example, we show that the second-order moment of the cluster size ${S}_2=\langle\sum\limits_{k}C_{k}^{2}\rangle_{\alpha}/N$ is equal to the susceptibility $\chi = \langle M^2 \rangle_\alpha/N$ of the AIM configurations, where $C_k$ refers to the size of the $k$-th cluster and
 $N$ represents the the total number of lattice sites. 
Using this relationship, the susceptibility, which has a divergent behavior characterized by the magnetic exponents near the critical point, remains unchanged using the AIM method.

To prove $\chi = S_2$, one can express the magnetization for the AIM configurations as,
	\begin{equation}
		M=\sum\limits_{i=1}^{N}s_{i}=\sum\limits_{k=1}^{N_C}C_{k}\tilde{s}_{k},
	\end{equation}
	where $\tilde{s}_k$ is the spin status ($\pm1$) of the $k$-th cluster. Then,
	the squared magnetization reads

 \begin{align}
 	\langle{M^{2}}\rangle_{\alpha}&=\langle\sum\limits_{k}\sum\limits_{k'}C_{k}C_{k'}\tilde{s}_{k}\tilde{s}_{k'}\rangle_{\alpha}\nonumber\\
&=\langle\sum\limits_{k}C_{k}^{2}\tilde{s}_{k}^{2}\rangle_{\alpha}+\langle \sum\limits_{k}\sum\limits_{k'\neq k}C_{k}C_{k'}\tilde{s}_{k}\tilde{s}_{k'}\rangle_{\alpha}\nonumber\\
	&=\langle\sum\limits_{k}C_{k}^{2}\rangle_{\alpha}\nonumber\ .
 \end{align}

 During the derivation, 
$
 \langle\sum\limits_{k}\sum\limits_{k'\neq k}C_{k}C_{k'}\tilde{s}_{k}\tilde{s}_{k'}\rangle_{\alpha}=0,
$
is used because the auxiliary Ising spins on different clusters are independent. The above derivation based on the site percolation models is applicable to any spatial dimensions and lattice (graph) structures. It can be readily generalized to the bond percolation models by only changing the way to construct percolation clusters.

\subsection{Learning performance for the percolation model }
\label{subsection:percolation--learning_the_phase_transition_of_percolation}

We classify the original configurations of the percolation and the AIM configurations using the confusion scheme. The main idea of the confusion scheme is to obtain the performances of neural networks trained with data that are marked with trial labels, and then identify the phase transition based on those performances. Details of the confusion scheme and the network structure used in our training can be found in Appendix~\ref{appendix:confusion_scheme}.

In Fig.~\ref{fig:all_in}, we compare the performance curves with and without using AIM for site and bond percolation on various lattices. A clear difference in the performance curve can be seen between the lower row the upper row, trained with and without AIM, respectively. In all the four percolation models we test, the performance curves of the original configurations are V-shaped, which means that the neural network fails to detect any phase transition features~\cite{confusion}. However, clear W-shaped performance curves emerge for networks trained on the AIM configurations, indicating that the percolation threshold is detected from the data. Since the confusion scheme is an unsupervised method, this result shows unsupervised machine learning is able to locate the percolation threshold based on AIM.

To verify that the detected phase transition is exactly the percolation transition, we compare the peak position of the W-shaped performance curve and the corresponding percolation threshold for the models. In Fig.~\ref{fig:all_in}(a2), the peak locates at the percolation threshold for the square-lattice site percolation $p_{c} = 0.59274605079210(2)$~\cite{Jacobsen_2015,PhysRevE.78.031136} within one discrete unit $\Delta p = 0.02$. This is also true for the site percolation model on the triangular lattice with $p_{c} = 1/2$~\cite{doi:10.1063/1.1704215}, the hexagonal lattice with $p_{c} = 0.697040230(5) $~\cite{threshold2,PhysRevE.78.031136} and the bond percolation model on the square lattice with $p_{c} =1/2$~\cite{kesten1980critical}, as shown in Figs.~\ref{fig:all_in} (b2)-(d2), respectively. We further train the neural networks on data generated in different sizes $L\times L$. The results in Fig.~\ref{fig:all_in} show that the peak positions converge to the percolation thresholds as the size increases, and thus we conclude that the phase transition detected by the confusion scheme from the auxiliary Ising spin configurations is exactly the percolation transition. 

\begin{figure}
 \centering
 \includegraphics[width=0.45\textwidth]{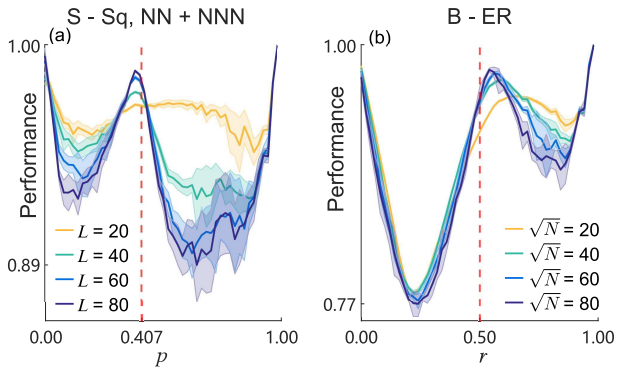}
 \caption{The confusion performance curve for percolation on graphs. The shaded area denotes the $3\sigma$ errorbar. The curve of (a) the site percolation on the square lattice with nearest and next-nearest-neighbouring bonds (S-Sq, NN+NNN). (b) ER Network bond percolation (B-ER), and the graphs have $N$ nodes and $rN$ edges. The periodic boundary condition is applied to the square lattice in (a).}
 \label{fig:more_all_1}
\end{figure}

We further demonstrate that the percolation transition on more complex network structures can be detected in an unsupervised way after auxiliary Ising mapping.
Figure~\ref{fig:more_all_1} shows the performance curve using AIM on (a) the site percolation on the square lattice with nearest and next-nearest-neighbouring bonds and (b) the bond percolation on the Erd\"os–R\'enyi (B-ER) network~\cite{erdos1960evolution}. A clear W-shape can be seen on the performance curves, and the peaks also locate near the percolation thresholds 0.407(2)~\cite{8neighbor,ActaPhysPB} and $1/2$~\cite{erdos1960evolution}, respectively. The detailed definition of B-ER percolation can be found in Appendix~\ref{appendix:ER_network}.

In Figs.~\ref{fig:all_in} and \ref{fig:more_all_1}, all peaks in the W-shaped graphs are located below the threshold except for the B-ER model. These deviations are non-universal.

For the B-ER model with $N$ sites, the pseudocritical point $r_c(N)$ estimated deviates from the threshold $r_c(\infty)$ and it can be expressed as:
  \begin{equation}
    r_c(N) =r_c(\infty) + a_{\widetilde{\nu}} N^{-1/\widetilde{\nu}},
    \label{eq:rc}
  \end{equation}
 where the critical exponent $\widetilde{\nu}$ is universal and describe the critical behavior of the system. However, $a_{\widetilde{\nu}}$ is not universal, and it depends on the model itself and the definition of the pseudo-critical point. For a given system, both $a_{\widetilde{\nu}}>0$ and $a_{\widetilde{\nu}}<0$ are possible, and  $r_c(N)$ can be greater or less than  $r_c(\infty)$. 
  The sign of $a_{\widetilde{\nu}}$ might be the result of many resonances and complexities when the confusion method is used.
  Another similar example is the three-dimensional Ising model, for which the different signs of $a$ emerge when using the magnetization and Binder ratio~\cite{landau}.

%We performed finite scaling analysis for the estimated percolation threshold of Erd\"os–R\'enyi model in Fig.~\ref{fig:more_all_1}, the result is shown in Fig.~\ref{fig:rc}.
\begin{figure}
 \centering
 \includegraphics[width = 0.40\textwidth]{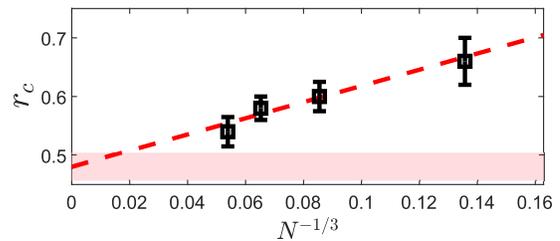}
  \caption{{Finite-size scaling analysis of the B-ER model. The pink area refers to the fitted percolation critical point $r_c(N\to +\infty) = 0.48(2)$ consistent with the theoretical value of $0.5$ within the error bar.}}
   \label{fig:rc}
\end{figure}

%   For the ``slow approach'' in the original manuscript, it is because the used of number of sites $N$ rather than the linear scale $L$ in ER model, the relationship between $N$ and $L$ is $N=L^2$ in our simulation. To reduce the confusion, we changed the label of FIG.5 in the manuscript from $N$ to $\sqrt{N}$ as the following:

\textcolor{blue}{
%Further more, thelocation of the maximum of the performance curve as a function of $N$ is studied, and $r_c(\infty)$ is obtained. 
Finite-size scaling analysis of critical point for the B-ER model 
is also performed. 
In Fig.~\ref{fig:rc}, by bringing $\widetilde{\nu}=3$ into Eq.~(\ref{eq:rc}),  the data of $r_c$
versus $N^{-1/3}$ is ploted and $r_c(\infty)$ is fitted to be 0.48(2), which is 
consistent with its theroetical value of 0.5 within the error bar.}

%With for the B-ER model~\cite{stauffer2018introduction} and the finite scaling analysis of $r_c$ is performed.}

\subsection{Correlated percolation: the Potts and XY models}

\label{section:Generalized_Percolation_potts_and_XY_model}

In this section, we study the correlated percolation transition in the Potts model and the XY model using the AIM method.

The Hamiltonians of both models are:
\begin{equation} 
 \left\{
 \begin{aligned}
 H_{Potts}&=-J\sum\limits_{\langle i,j\rangle} \delta_{\sigma_i \sigma_j}\\
 H_{XY}&=-J\sum\limits_{\langle i,j\rangle}\myvect{s_{i}}\cdot\myvect{s_{j}}
 \end{aligned}
 \right.\ ,
 \end {equation}
% \textcolor{blue}{Changed: $S\to s$}
where the Potts variable takes values $\sigma=1,2,...,q$ and the XY spin $\myvect{s_i}$ is unit vector with two real components.
For both models, the ferromagnetic coupling strength is fixed with $J>0$ and the summation is performed over nearest-neighbor pairs $ \langle i,j\rangle $ of sites. Both of the Potts ~\cite{tan2020comprehensive,li2018applications,potts} and the XY models~~\cite{wessel, cf_qlr,Zhang_W,PhysRevResearch.3.013074,df1} have been studied using machine learning methods extensively.

The Potts model can be seen as a correlated percolation model in the Fortuin-Kasteleyn (FK) representation, which is generated by connecting the nearest neighbors with the probability $p_{ij}=1-e^{-\beta J \delta_{\sigma_i,\sigma_j}}$ ~\cite{fk}. The inverse temperature $\beta$ is defined as $\beta = 1/T$. It is shown that the correlation function of the corresponding correlated percolation model is proportional to the correlation function in the original Potts model with a fixed coefficient depending on $q$, due to the similarity of the mapped partition functions~\cite{fk}. This proportionality leads directly to a consistent critical behavior, which relates the thermal phase transition of the Potts model to the percolation transition in the FK representation~\cite{fk}.

A similar correlated percolation model existed in the FK representation after being projected in one chosen direction for the XY model~\cite{perco-xy1}. The projected XY model can be obtained by choosing a randomly oriented Cartesian frame of reference $(x,y)$ in the spin space and projecting all spins along the $x$ and $y$ axes: $\myvect{s}_i = s^x_i \hat{x} + s^y_i\hat{y}$.%\textcolor{blue}{$S\to s$}
~The XY model then turns out to be two Ising-like models: $H_{XY} = H_{x} + H_{y}$. In one Ising-like projected XY model, such as that one projected on $\hat{x}$, the correlated percolation model in the FK representation can be built by connecting nearest sites with the probability $p_{ij}=\max(0,1-e^{-2\beta J s^x_i s^x_j})$~\cite{perco-xy1,PhysRevLett.58.86}. Due to a similar reason as that in the Potts model, the correlation function restricted on the direction $\hat{x}$ of the XY model is preserved. Therefore the phase transition in the XY model can be inferred from the percolation transition of the correlated percolation model.
 
\begin{figure}
 \centering
 \includegraphics[width = 0.45\textwidth]{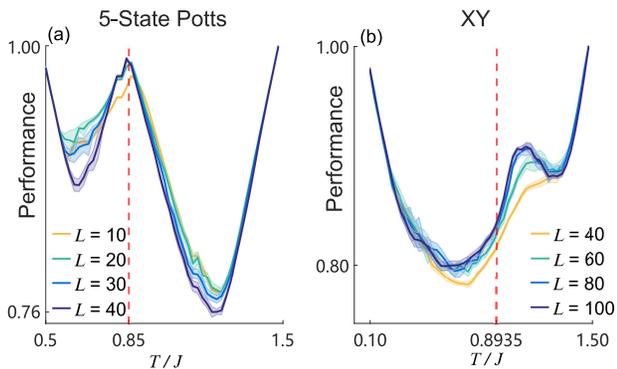}
 \caption{The confusion performance curves for generalized percolation. The shaded area denotes the $3\sigma$ errorbar. The curves of the (a) $q=5$ Potts model and (b) the XY model. The red line denotes the percolation threshold. The periodic boundary condition is applied to all lattices.}
 \label{fig:more_all_2}
\end{figure}

For the correlated percolation representation generated based on the Potts and the XY models, the AIM configurations can be obtained by randomly assigning $\pm 1$ to different clusters. The previous proof also applies here, that, the correlation functions of the correlated percolation and the auxiliary Ising spin model are equal.

Then the confusion method can be applied to detect the phase transition of the correlated percolation model in an unsupervised way. In Figs.~\ref{fig:more_all_2} (a) and (b), the performance curves by using the AIM configurations of the $q=5$ Potts and the XY models as training data are shown, respectively. Clear W-shaped performance curves emerge for the Potts model, indicating that the confusion method detects a phase transition. Furthermore, the peak position of the performance curve is consistent with the critical point $T_c/J = 1/\ln(1+\sqrt{5})=0.852$~\cite{Hintermann1978} within one numerical discrete step $\Delta T = 0.02J$. 

As for the XY model, W-shaped performance curves also show up at sizes $L=60, 80, 100$ and the peaks of the performance curves are about $T/J\approx 1.04$, which is approximate to the result without using the AIM transformation~\cite{wessel,cf_qlr}. Although the peak position has a tendency to become closer to the correct BKT critical point as the system size increases, we still can not claim a clear relation between the peak and the BKT critical point $T/J \approx 0.8935(1)$~\cite{BKT,Hsieh_2013}. We suggest that this slow-converging peak position is probably a result of the logarithmic finite-size correction of stiffness and linear size $L$ at the BKT phase transition point~\cite{Kosterlitz_1974,PhysRevB.37.5986,Hsieh_2013}. The deviations from the estimated values of the phase transition points in Refs.~\cite{wessel,cf_qlr} probably come from the same reason.

\section{Identification of Universalities}

The study of universality and the critical exponent is an important task in statistical physics~\cite{universality}. Machine learning, as an emerging approach, has been applied recently to extract universality and critical exponent~\cite{Critical_Exponent1, Giannetti}. In the two dimensional $q$-state Potts model, a second-order transition occurs at $q\leq 4$ and a first-order phase transition for otherwise ~\cite{baxter1973potts,baxter1978triangular}. For 2-state Potts and other Potts models, the difference in universalities leads to different correlated percolation configurations in the FK representation at the phase transition point, respectively. This difference is difficult to distinguish by conventional data analysis methods.

The AIM method is a correlation-preserved transformation, therefore the universality and scaling behavior remain unchanged after the mapping. In this section, we demonstrate that the universality differences in different Potts model can be detected with AIM.

By converting the Potts model to a correlated percolation model, we generate the AIM configurations of the $q$=$\{$2,3,4,5,6,7,8,9,10$\}$ state Potts models at their respective critical points at different system sizes. %For each $q$-state Potts model, we composed a training set with equal number $q$-state Potts configuration and 2-state Potts configuration labeled with 0 and 1 correspondingly and then trained the neural network supervised. Finally, we apply the trained neural network to a test set with an equal number of $q$-state Potts configuration and 2-state Potts configuration to obtain the binary classification accuracy.}
Then, we try to distinguish the $q_0 = 2$-state Potts models and other $q$-state Potts models in turn according to their configurations at the phase transition points by supervised machine learning.
The training and test dataset contains both types of configurations with correct labels. The configurations of the $q_0$-state Potts are labeled with $1$ or the positive (P) class, and the ones of $q$-state Potts model are labeled with $0$ or the negative (N) class. The neural network is then trained in the way described in Appendix.~\ref{subsection:appendix--Binary_Classification_for_Universal_Class_Classification}. Finally, we apply the trained neural network to the test dataset to predict the labels and calculate the binary classification accuracy, which is defined to be the ratio of the number of correctly predicted samples to the total number of samples~\cite{goodfellow2016deep}:

 \begin{equation}
 \text{Accuracy} = \frac{\text{TP+TN}}{\text{TP+FP+TN+FN}}\ \ ,
 \label{eq:binary_accuracy}
\end{equation}

where true-positive (TP) is the number of samples for which the model correctly predicts the ``P" class, and the other three cases true-negative (TN), false-positive (FP) and false-negative (FP) are defined in a simialr way.

\begin{figure}[t]
 \centering
 \includegraphics[width=0.30\textwidth]{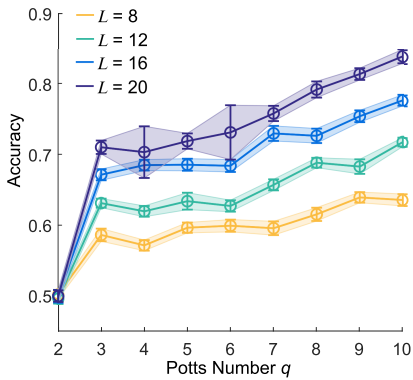}
 \caption{{Binary classification accuracy of 2-state and other $q$-state Potts model for different system sizes. The data input to the neural network are all auxiliary Ising configurations mapped from the Potts model at the corresponding phase transition point.}}
 \label{fig:phase_diff}
\end{figure}

Figure~\ref{fig:phase_diff} shows the binary classification accuracy. Here we should briefly explain why the classification accuracy is 0.5 for $q = q_0 =2$, i.e., both sets of samples involved in the classification are the 2-state Potts model. The two sets of samples are essentially indistinguishable, but are given different labels as N and P, respectively. Therefore the neural network used for binary classification learns nothing. The equation TP=FP=TN=FN holds and the value of Eq. (\ref{eq:binary_accuracy}) becomes 0.5 if both datasets have the same number of samples.  

For general values of $q_0$, two features can be observed: (I), as the Potts number $q$ increases, the binary classification accuracy increases. This is because the $q$-state Potts model tends to have a more drastic phase transition as $q$ increases. (II), the binary classification accuracy increases as the size of the system increases, which indicates that the accuracy acquired by the neural network is probably 100\% in the thermodynamic limit.

\label{section:Identification_of_universal_classes}

In summary, the above results imply that the neural network can distinguish different universalities from the configurations obtained after the AIM transformation.

\section{Summary and Discussion}
\label{section:Summary_and_Discussion}

In this paper, we propose the auxiliary Ising mapping method for machine learning to detect both uncorrelated and correlated percolation transition in an unsupervised way. To demonstrate that the robustness of this method, we have validated it under site percolation model, bond percolation model, Potts model, and XY model on different lattices. The results show that the confusion scheme, which failed in the percolation model, could learn the geometric phase transition correctly with our mapping method.
For the XY model, the observed deviated peak position probably come from a logarithmic finite-size correction. Furthermore, we also show that the universalities of the Potts model with different values of $q$ can be distinguished by neural networks with the AIM method.

There are many ways in which the state of a physical system can be represented as a dataset, and applying machine learning to a more feature-specific representation can make machine learning more effective. Auxiliary variables, e.g., the Ising-like spins, can be one of such representations. This approach might advance the application of machine learning in statistical and condensed-matter physics.

\section*{Acknowledgments}
\label{section:Acknowledgements}
We thank Heyang Ma and Guimin Lin for discussions. 
 Y. D is supported by the National Natural Science Foundation of China (under Grant No. 11625522),
the Science and Technology Committee of Shanghai (under grant No. 20DZ2210100), the National Key R\&D
Program of China (under Grant No. 2018YFA0306501).
W. Z is supported by the
 open project KQI201 from Key Laboratory of Quantum Information, University of Science and Technology of China, Chinese Academy of Sciences, and the open project by Hefei National Laboratory for Physical Sciences at the Microscale

\appendix

 \section{The structure of Network}
 In both the confusion scheme and the universality classification, we use the binary classification neural networks. The binary classification neural network is designed for the classification of the samples in a supervised way, which is generally composed of three main parts: the input layer, the hidden layer(s), and the output layer. The input layer, which has the same dimension as the input data, can be both a fully connected layer or convolution layer. For the hidden part, there can be more than one layer, and a composed structure can be utilized by combining the convolution layer and the fully connected layers. For the output layer in a binary classification neural network, there are generally two choices: one output form and two output forms, which are equivalent to each other. In our work, we used the previous one. For a binary classification task, all the samples are pre-labeled with $0$ or $1$, which refers to the two different classes. The samples are then separated into a training set and a test. Gradient descent algorithm is used to train the neural network out of the training set in other to teach the neural network to find the hidden difference within the $0$ and $1$ data. The loss function we used in binary classification is the binary cross-entropy, which is defined as~\cite{goodfellow2016deep}:
 \begin{equation}
 L = \frac{1}{N}\sum_i -\left(
 y_i \log\left(p_i\right) + \left(1-y_i\right) \log\left(1-p_i\right)
 \right)\ ,
 \end{equation}
 where $N$ is the number of the data in the training set, $y_i$ is the label ($0$ or $1$) for the $i$-th sample and $p_i$ is the output of the output neuron, which refers to the probability that the $i$-th sample is predicted with label $1$. It can be seen that, by minimizing the binary cross-entropy, the neural network is approaching a better classification out of the training set. To quantify the training result, we further apply the neural network we trained onto the test set and give the predicted label $y_i^\prime$ for the $i$-th sample according to the output neural: if the output is larger than $0.5$, then $y_i^\prime = 1$, otherwise, $y_i^\prime = 0$. We quantify the performance of the classification network by the binary classification accuracy defined in Eq.(~\ref{eq:binary_accuracy}). In the confusion scheme, the binary classification accuracy obtained by training the network using trial labels is called the confusion performance.

 \label{section:appendix--structure_of_network}
 \begin{figure*}
 \centering
 \includegraphics[width=0.8\textwidth]{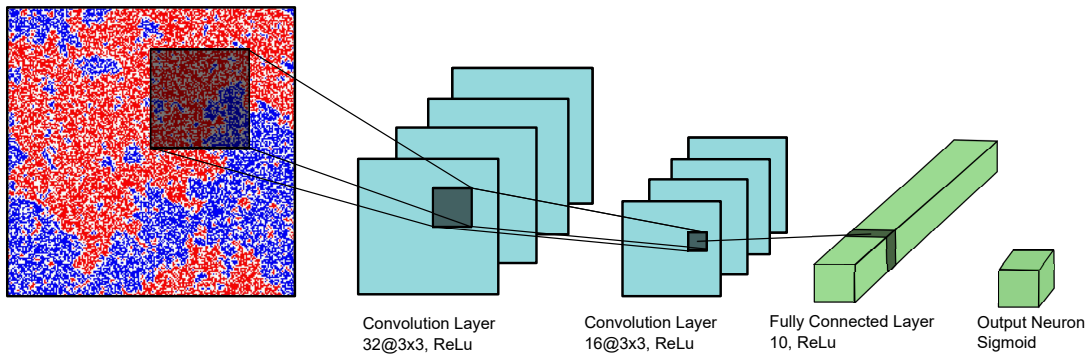}
 \caption{
 The structure of the binary classification network is used to classify the different universalities of the Potts model. Layer One: Convolutionlayer, 32 channels, the kernel size is 3×3, activated by ReLu. Layer Two: Convolution layer, 16 channels, the kernel size is 3×3, activated by ReLu. Layer Three: Fully connected layer, 10 neurons, activated by ReLu. Output layer: one output neuron activated by Sigmoid function. }
 \label{fig:cnn}
\end{figure*}

 \subsection{Binary Classification for Confusion Scheme}
 \label{subsection:appendix--Binary_Classification_for_confusion}
 The structure of this neural network is shown in Fig.~\ref{fig:first_diagram}. The neural network we use has one input layer, one hidden layer, 10 neurons in the hidden layer, and the ReLu function is chosen as the activation function. An L2-regularization with a rate of 0.001 to avoid over-fitting. The output neuron uses Sigmoid as the activation function to obtain the output in the interval $[0,1]$ to indicate the classification accuracy. We randomly select 40,000 samples out of 60,000 as the training set and the rest as the test set in each confusion training. In the training process, we set the batch size of the training set to 512, use the RmsProp algorithm for adaptive gradient descent training, train epoch to 30, and finally use the cross-entropy as the loss function.
 
 \subsection{Binary Classification for universality Classification}
 \label{subsection:appendix--Binary_Classification_for_Universal_Class_Classification}
 
 The structure of the binary classification network used to classify different universalities of Potts model is shown in Fig.~\ref{fig:cnn}. The first two layers are convolution networks, the kernel size is $3\times 3$, the first layer has 32 channels while the second layer has 16 channels. A fully connected layer with ten neurons is placed behind the convolutional layer. Rectified linear unit (ReLu) is chosen as the activation function for the first three layers. The output neuron is activated by Sigmoid function. In order to avoid over-fitting, L2 regularization is used in the third layer (fully connected layer with 10 neurons), the L2 regularization rate is 0.001. 
 
 To obtain the data used in Fig.~\ref{fig:phase_diff}, the training set consists of 8000 samples, while the test set has 2000 samples. The number of training epoch is 10, and the batchsize is chosen to be 32 in each epoch.
 
 \section{Confusion scheme}
 \label{appendix:confusion_scheme}
\label{Confusion}

 The confusion scheme is invented to detect phase transition and locates the phase transition point without knowing any prior knowledge of the phases. The main technique used in the confusion scheme is the use of trial labels. For a phase transition controlled by one variable, such as the 2D Ising model without external magnetic fields controlled by temperature $T$, the configurations can be classified into two different classes: the ferromagnetic phase $T<T_c$ and the disordered phase $T>T_c$. Therefore, if we have to separate the data with different $T$ into two classes, the separation according to $T_c$ will result in the maximum differences between the two classes, therefore generally leads to a higher binary classification accuracy obtained by the neural network. The main idea of the confusion scheme is to find such accuracy peaks, or rather, performance peaks, which help to detect phase transitions and assign transition points. The algorithm for the confusion scheme is shown as the following:

 (1) Prepare a dataset for machine learning. 
 Taking the site-percolation model as an example, we divide the parameter range $[p_{min},p_{max}]$ into $N_{p}$ parameters $\{p_i\}$ and generate $N_{s}$ samples of both the original configurations and the auxiliary Ising spin configurations at each parameter $p_i$. This gives us $N=N_{p}\times N_{s}$ configurations.
 
(2) Make the trial labels and perform training and testing.
 The $p_i$ is chosen as the trial critical point for generating trial labels, and all samples generated with $p_j<p_i$ are assigned label $0$, and the rest are assigned label $1$.
A binary classification neural network is then trained to classify the samples while recording the accuracy $Acc(p_i)$ or called performance.

(3) Repeat step 2 with all trial critical points $\{p_i\}$ in the range $[p_{min},p_{max}]$, and obtain the performance curve. 

For the trial critical locates at the lower (or upper) boundary for the parameter regime, all trial labels are $0$ (or $1$), which results in a perfect classification with $100\%$ accuracy. Therefore, if there are no phase transitions in the parameter investigated parameter range, or the difference between two phases is still too difficult for the neural network to capture, then a V-shaped performance curve can be expected.

From the reasoning above, we also know that the local maximum of performance will be obtained if the trial labels exactly refer to different phases. In that case, the performance curve would have several local maxima that separate different phases, therefore, allocate the phase transition point. For a special case that the system has only two distinguished phases, such as the site percolation model on the square lattice, then the performance curve would be a W-shaped curve, and the peak locates exactly at the phase transition point.

 \section{ER network}
 \label{appendix:ER_network}
\label{ER}

Erd\"os–R\'enyi model is initially proposed by Paul Erd\"os and Alfr\'ed R\'enyi for generating random graphs~\cite{erdos1960evolution}.

The ER network consists of $N$ isolated vertices and $rN$ edges are added randomly to connect those vertices. By varying the percolation parameter
$r$, the system undergoes a second-order phase transition at $r=r_c=1/2$.
For $r<r_c$, the ER network is disjoint with small clusters; for $r>r_c$, the percolated cluster emerges.

 \bibliography{referencesnew}

\begin{thebibliography}{56}%
\makeatletter
\providecommand \@ifxundefined [1]{%
 \@ifx{#1\undefined}
}%
\providecommand \@ifnum [1]{%
 \ifnum #1\expandafter \@firstoftwo
 \else \expandafter \@secondoftwo
 \fi
}%
\providecommand \@ifx [1]{%
 \ifx #1\expandafter \@firstoftwo
 \else \expandafter \@secondoftwo
 \fi
}%
\providecommand \natexlab [1]{#1}%
\providecommand \enquote  [1]{``#1''}%
\providecommand \bibnamefont  [1]{#1}%
\providecommand \bibfnamefont [1]{#1}%
\providecommand \citenamefont [1]{#1}%
\providecommand \href@noop [0]{\@secondoftwo}%
\providecommand \href [0]{\begingroup \@sanitize@url \@href}%
\providecommand \@href[1]{\@@startlink{#1}\@@href}%
\providecommand \@@href[1]{\endgroup#1\@@endlink}%
\providecommand \@sanitize@url [0]{\catcode `\\12\catcode `\$12\catcode
  `\&12\catcode `\#12\catcode `\^12\catcode `\_12\catcode `\%12\relax}%
\providecommand \@@startlink[1]{}%
\providecommand \@@endlink[0]{}%
\providecommand \url  [0]{\begingroup\@sanitize@url \@url }%
\providecommand \@url [1]{\endgroup\@href {#1}{\urlprefix }}%
\providecommand \urlprefix  [0]{URL }%
\providecommand \Eprint [0]{\href }%
\providecommand \doibase [0]{https://doi.org/}%
\providecommand \selectlanguage [0]{\@gobble}%
\providecommand \bibinfo  [0]{\@secondoftwo}%
\providecommand \bibfield  [0]{\@secondoftwo}%
\providecommand \translation [1]{[#1]}%
\providecommand \BibitemOpen [0]{}%
\providecommand \bibitemStop [0]{}%
\providecommand \bibitemNoStop [0]{.\EOS\space}%
\providecommand \EOS [0]{\spacefactor3000\relax}%
\providecommand \BibitemShut  [1]{\csname bibitem#1\endcsname}%
\let\auto@bib@innerbib\@empty
%</preamble>
\bibitem [{\citenamefont {Broadbent}\ and\ \citenamefont
  {Hammersley}(1957)}]{bm}%
  \BibitemOpen
  \bibfield  {author} {\bibinfo {author} {\bibfnamefont {S.~R.}\ \bibnamefont
  {Broadbent}}\ and\ \bibinfo {author} {\bibfnamefont {J.~M.}\ \bibnamefont
  {Hammersley}},\ }\bibfield  {title} {\bibinfo {title} {{Percolation
  processes: I. Crystals and mazes}},\ }in\ \href
  {https://www.semanticscholar.org/paper/Percolation-processes.-I.-Crystals-and-Mazes-Broadbent-Hammersley/55caf92bd5d58d6c2f5593ceccb2fd1916fd5340}
  {\emph {\bibinfo {booktitle} {Math. Proc. Camb. Philos. Soc.}}},\
  Vol.~\bibinfo {volume} {53}\ (\bibinfo  {publisher} {Cambridge University
  Press},\ \bibinfo {year} {1957})\ pp.\ \bibinfo {pages}
  {629--641}\BibitemShut {NoStop}%
\bibitem [{\citenamefont {Hunt}\ \emph {et~al.}(2014)\citenamefont {Hunt},
  \citenamefont {Ewing},\ and\ \citenamefont {Ghanbarian}}]{5}%
  \BibitemOpen
  \bibfield  {author} {\bibinfo {author} {\bibfnamefont {A.}~\bibnamefont
  {Hunt}}, \bibinfo {author} {\bibfnamefont {R.}~\bibnamefont {Ewing}},\ and\
  \bibinfo {author} {\bibfnamefont {B.}~\bibnamefont {Ghanbarian}},\ }\href
  {https://link.springer.com/book/10.1007/978-3-319-03771-4} {\emph {\bibinfo
  {title} {{Percolation theory for flow in porous media}}}},\ Vol.\ \bibinfo
  {volume} {880}\ (\bibinfo  {publisher} {Springer},\ \bibinfo {year}
  {2014})\BibitemShut {NoStop}%
\bibitem [{\citenamefont {Newman}(2002)}]{6}%
  \BibitemOpen
  \bibfield  {author} {\bibinfo {author} {\bibfnamefont {M.~E.~J.}\
  \bibnamefont {Newman}},\ }\bibfield  {title} {\bibinfo {title} {Spread of
  epidemic disease on networks},\ }\href
  {https://doi.org/10.1103/PhysRevE.66.016128} {\bibfield  {journal} {\bibinfo
  {journal} {Phys. Rev. E}\ }\textbf {\bibinfo {volume} {66}},\ \bibinfo
  {pages} {16128} (\bibinfo {year} {2002})}\BibitemShut {NoStop}%
\bibitem [{\citenamefont {Duffie}\ and\ \citenamefont
  {Manso}(2007)}]{duffie2007information}%
  \BibitemOpen
  \bibfield  {author} {\bibinfo {author} {\bibfnamefont {D.}~\bibnamefont
  {Duffie}}\ and\ \bibinfo {author} {\bibfnamefont {G.}~\bibnamefont {Manso}},\
  }\bibfield  {title} {\bibinfo {title} {Information percolation in large
  markets},\ }\href {https://doi.org/10.1257/aer.97.2.203} {\bibfield
  {journal} {\bibinfo  {journal} {Am. Econ. Rev.}\ }\textbf {\bibinfo {volume}
  {97}},\ \bibinfo {pages} {203} (\bibinfo {year} {2007})}\BibitemShut
  {NoStop}%
\bibitem [{\citenamefont {Men'shikov}\ \emph {et~al.}(1988)\citenamefont
  {Men'shikov}, \citenamefont {Molchanov},\ and\ \citenamefont
  {Sidorenko}}]{1986Percolation}%
  \BibitemOpen
  \bibfield  {author} {\bibinfo {author} {\bibfnamefont {M.~V.}\ \bibnamefont
  {Men'shikov}}, \bibinfo {author} {\bibfnamefont {S.~A.}\ \bibnamefont
  {Molchanov}},\ and\ \bibinfo {author} {\bibfnamefont {A.~F.}\ \bibnamefont
  {Sidorenko}},\ }\bibfield  {title} {\bibinfo {title} {Percolation theory and
  some applications},\ }\href
  {https://link.springer.com/article/10.1007/BF01095508} {\bibfield  {journal}
  {\bibinfo  {journal} {J. Sov. Math.}\ }\textbf {\bibinfo {volume} {42}},\
  \bibinfo {pages} {1766} (\bibinfo {year} {1988})}\BibitemShut {NoStop}%
\bibitem [{\citenamefont {Cohen}\ \emph {et~al.}(2011)\citenamefont {Cohen},
  \citenamefont {Erez}, \citenamefont {Havlinl}, \citenamefont {Newman},
  \citenamefont {Barab{\'{a}}si},\ and\ \citenamefont {Watts}}]{8}%
  \BibitemOpen
  \bibfield  {author} {\bibinfo {author} {\bibfnamefont {R.}~\bibnamefont
  {Cohen}}, \bibinfo {author} {\bibfnamefont {K.}~\bibnamefont {Erez}},
  \bibinfo {author} {\bibfnamefont {S.}~\bibnamefont {Havlinl}}, \bibinfo
  {author} {\bibfnamefont {M.}~\bibnamefont {Newman}}, \bibinfo {author}
  {\bibfnamefont {A.-L.}\ \bibnamefont {Barab{\'{a}}si}},\ and\ \bibinfo
  {author} {\bibfnamefont {D.~J.}\ \bibnamefont {Watts}},\ }\bibfield  {title}
  {\bibinfo {title} {Resilience of the internet to random breakdowns},\ }in\
  \href {http://tuprints.ulb.tu-darmstadt.de/1286/1/diss_dyn_network.pdf}
  {\emph {\bibinfo {booktitle} {The Structure and Dynamics of Networks}}}\
  (\bibinfo  {publisher} {Princeton University Press},\ \bibinfo {year}
  {2011})\ pp.\ \bibinfo {pages} {507--509}\BibitemShut {NoStop}%
\bibitem [{\citenamefont {Callaway}\ \emph {et~al.}(2000)\citenamefont
  {Callaway}, \citenamefont {Newman}, \citenamefont {Strogatz},\ and\
  \citenamefont {Watts}}]{9}%
  \BibitemOpen
  \bibfield  {author} {\bibinfo {author} {\bibfnamefont {D.~S.}\ \bibnamefont
  {Callaway}}, \bibinfo {author} {\bibfnamefont {M.~E.~J.}\ \bibnamefont
  {Newman}}, \bibinfo {author} {\bibfnamefont {S.~H.}\ \bibnamefont
  {Strogatz}},\ and\ \bibinfo {author} {\bibfnamefont {D.~J.}\ \bibnamefont
  {Watts}},\ }\bibfield  {title} {\bibinfo {title} {Network robustness and
  fragility: Percolation on random graphs},\ }\href
  {https://doi.org/10.1103/PhysRevLett.85.5468} {\bibfield  {journal} {\bibinfo
   {journal} {Phys. Rev. Lett.}\ }\textbf {\bibinfo {volume} {85}},\ \bibinfo
  {pages} {5468} (\bibinfo {year} {2000})}\BibitemShut {NoStop}%
\bibitem [{\citenamefont {Achlioptas}\ \emph {et~al.}(2009)\citenamefont
  {Achlioptas}, \citenamefont {D'Souza},\ and\ \citenamefont
  {Spencer}}]{0Explosive}%
  \BibitemOpen
  \bibfield  {author} {\bibinfo {author} {\bibfnamefont {D.}~\bibnamefont
  {Achlioptas}}, \bibinfo {author} {\bibfnamefont {R.~M.}\ \bibnamefont
  {D'Souza}},\ and\ \bibinfo {author} {\bibfnamefont {J.}~\bibnamefont
  {Spencer}},\ }\bibfield  {title} {\bibinfo {title} {Explosive percolation in
  random networks},\ }\href
  {https://www.science.org/lookup/doi/10.1126/science.1167782} {\bibfield
  {journal} {\bibinfo  {journal} {Science}\ }\textbf {\bibinfo {volume}
  {323}},\ \bibinfo {pages} {1453} (\bibinfo {year} {2009})}\BibitemShut
  {NoStop}%
\bibitem [{\citenamefont {Li}\ \emph {et~al.}(2021)\citenamefont {Li},
  \citenamefont {Liu}, \citenamefont {Lü}, \citenamefont {Hu}, \citenamefont
  {Xu},\ and\ \citenamefont {Zhang}}]{2021Percolation}%
  \BibitemOpen
  \bibfield  {author} {\bibinfo {author} {\bibfnamefont {M.}~\bibnamefont
  {Li}}, \bibinfo {author} {\bibfnamefont {R.-R.}\ \bibnamefont {Liu}},
  \bibinfo {author} {\bibfnamefont {L.}~\bibnamefont {Lü}}, \bibinfo {author}
  {\bibfnamefont {M.-B.}\ \bibnamefont {Hu}}, \bibinfo {author} {\bibfnamefont
  {S.}~\bibnamefont {Xu}},\ and\ \bibinfo {author} {\bibfnamefont {Y.-C.}\
  \bibnamefont {Zhang}},\ }\bibfield  {title} {\bibinfo {title} {Percolation on
  complex networks: Theory and application},\ }\href
  {https://doi.org/10.1016/j.physrep.2020.12.003} {\bibfield  {journal}
  {\bibinfo  {journal} {Phys. Rep.}\ }\textbf {\bibinfo {volume} {907}},\
  \bibinfo {pages} {1} (\bibinfo {year} {2021})}\BibitemShut {NoStop}%
\bibitem [{\citenamefont {Fortuin}\ and\ \citenamefont {Kasteleyn}(1972)}]{fk}%
  \BibitemOpen
  \bibfield  {author} {\bibinfo {author} {\bibfnamefont {C.~M.}\ \bibnamefont
  {Fortuin}}\ and\ \bibinfo {author} {\bibfnamefont {P.~W.}\ \bibnamefont
  {Kasteleyn}},\ }\bibfield  {title} {\bibinfo {title} {{On the random-cluster
  model: I. Introduction and relation to other models}},\ }\href
  {https://doi.org/10.1016/0031-8914(72)90045-6} {\bibfield  {journal}
  {\bibinfo  {journal} {Physica}\ }\textbf {\bibinfo {volume} {57}},\ \bibinfo
  {pages} {536} (\bibinfo {year} {1972})}\BibitemShut {NoStop}%
\bibitem [{\citenamefont {Fortuin}(1972)}]{fk_two}%
  \BibitemOpen
  \bibfield  {author} {\bibinfo {author} {\bibfnamefont {C.}~\bibnamefont
  {Fortuin}},\ }\bibfield  {title} {\bibinfo {title} {On the random-cluster
  model ii. the percolation model},\ }\href
  {https://doi.org/https://doi.org/10.1016/0031-8914(72)90161-9} {\bibfield
  {journal} {\bibinfo  {journal} {Physica}\ }\textbf {\bibinfo {volume} {58}},\
  \bibinfo {pages} {393} (\bibinfo {year} {1972})}\BibitemShut {NoStop}%
\bibitem [{\citenamefont {Fisher}(1967)}]{fisher1967physics}%
  \BibitemOpen
  \bibfield  {author} {\bibinfo {author} {\bibfnamefont {M.~E.}\ \bibnamefont
  {Fisher}},\ }\bibfield  {title} {\bibinfo {title} {Magnetic critical point
  exponents—their interrelations and meaning},\ }\href
  {https://doi.org/10.1063/1.1709711} {\bibfield  {journal} {\bibinfo
  {journal} {J. Appl. Phys.}\ }\textbf {\bibinfo {volume} {38}},\ \bibinfo
  {pages} {981} (\bibinfo {year} {1967})}\BibitemShut {NoStop}%
\bibitem [{\citenamefont {Coniglio}\ and\ \citenamefont
  {Klein}(1980)}]{coniglio1980clusters}%
  \BibitemOpen
  \bibfield  {author} {\bibinfo {author} {\bibfnamefont {A.}~\bibnamefont
  {Coniglio}}\ and\ \bibinfo {author} {\bibfnamefont {W.}~\bibnamefont
  {Klein}},\ }\bibfield  {title} {\bibinfo {title} {Clusters and {Ising}
  critical droplets: a renormalisation group approach},\ }\href
  {https://doi.org/10.1088/0305-4470/13/8/025} {\bibfield  {journal} {\bibinfo
  {journal} {J. Phys. A Math. Theor.}\ }\textbf {\bibinfo {volume} {13}},\
  \bibinfo {pages} {2775} (\bibinfo {year} {1980})}\BibitemShut {NoStop}%
\bibitem [{\citenamefont {Hu}(1984)}]{hu}%
  \BibitemOpen
  \bibfield  {author} {\bibinfo {author} {\bibfnamefont {C.-K.}\ \bibnamefont
  {Hu}},\ }\bibfield  {title} {\bibinfo {title} {Percolation, clusters, and
  phase transitions in spin models},\ }\href
  {https://doi.org/10.1103/PhysRevB.29.5103} {\bibfield  {journal} {\bibinfo
  {journal} {Phys. Rev. B}\ }\textbf {\bibinfo {volume} {29}},\ \bibinfo
  {pages} {5103} (\bibinfo {year} {1984})}\BibitemShut {NoStop}%
\bibitem [{\citenamefont {Swendsen}\ and\ \citenamefont
  {Wang}(1987)}]{PhysRevLett.58.86}%
  \BibitemOpen
  \bibfield  {author} {\bibinfo {author} {\bibfnamefont {R.~H.}\ \bibnamefont
  {Swendsen}}\ and\ \bibinfo {author} {\bibfnamefont {J.-S.}\ \bibnamefont
  {Wang}},\ }\bibfield  {title} {\bibinfo {title} {Nonuniversal critical
  dynamics in {Monte Carlo} simulations},\ }\href
  {https://doi.org/10.1103/PhysRevLett.58.86} {\bibfield  {journal} {\bibinfo
  {journal} {Phys. Rev. Lett.}\ }\textbf {\bibinfo {volume} {58}},\ \bibinfo
  {pages} {86} (\bibinfo {year} {1987})}\BibitemShut {NoStop}%
\bibitem [{\citenamefont {Carleo}\ \emph {et~al.}(2019)\citenamefont {Carleo},
  \citenamefont {Cirac}, \citenamefont {Cranmer}, \citenamefont {Daudet},
  \citenamefont {Schuld}, \citenamefont {Tishby}, \citenamefont
  {Vogt-Maranto},\ and\ \citenamefont {Zdeborov\'a}}]{review}%
  \BibitemOpen
  \bibfield  {author} {\bibinfo {author} {\bibfnamefont {G.}~\bibnamefont
  {Carleo}}, \bibinfo {author} {\bibfnamefont {I.}~\bibnamefont {Cirac}},
  \bibinfo {author} {\bibfnamefont {K.}~\bibnamefont {Cranmer}}, \bibinfo
  {author} {\bibfnamefont {L.}~\bibnamefont {Daudet}}, \bibinfo {author}
  {\bibfnamefont {M.}~\bibnamefont {Schuld}}, \bibinfo {author} {\bibfnamefont
  {N.}~\bibnamefont {Tishby}}, \bibinfo {author} {\bibfnamefont
  {L.}~\bibnamefont {Vogt-Maranto}},\ and\ \bibinfo {author} {\bibfnamefont
  {L.}~\bibnamefont {Zdeborov\'a}},\ }\bibfield  {title} {\bibinfo {title}
  {Machine learning and the physical sciences},\ }\href
  {https://doi.org/10.1103/RevModPhys.91.045002} {\bibfield  {journal}
  {\bibinfo  {journal} {Rev. Mod. Phys.}\ }\textbf {\bibinfo {volume} {91}},\
  \bibinfo {pages} {045002} (\bibinfo {year} {2019})}\BibitemShut {NoStop}%
\bibitem [{\citenamefont {Wang}(2016)}]{pca1}%
  \BibitemOpen
  \bibfield  {author} {\bibinfo {author} {\bibfnamefont {L.}~\bibnamefont
  {Wang}},\ }\bibfield  {title} {\bibinfo {title} {Discovering phase
  transitions with unsupervised learning},\ }\href
  {https://doi.org/10.1103/PhysRevB.94.195105} {\bibfield  {journal} {\bibinfo
  {journal} {Phys. Rev. B}\ }\textbf {\bibinfo {volume} {94}},\ \bibinfo
  {pages} {195105} (\bibinfo {year} {2016})}\BibitemShut {NoStop}%
\bibitem [{\citenamefont {Wetzel}(2017)}]{pca2}%
  \BibitemOpen
  \bibfield  {author} {\bibinfo {author} {\bibfnamefont {S.~J.}\ \bibnamefont
  {Wetzel}},\ }\bibfield  {title} {\bibinfo {title} {Unsupervised learning of
  phase transitions: From principal component analysis to variational
  autoencoders},\ }\href {https://doi.org/10.1103/PhysRevE.96.022140}
  {\bibfield  {journal} {\bibinfo  {journal} {Phys. Rev. E}\ }\textbf {\bibinfo
  {volume} {96}},\ \bibinfo {pages} {022140} (\bibinfo {year}
  {2017})}\BibitemShut {NoStop}%
\bibitem [{\citenamefont {Hu}\ \emph {et~al.}(2017)\citenamefont {Hu},
  \citenamefont {Singh},\ and\ \citenamefont {Scalettar}}]{pca3}%
  \BibitemOpen
  \bibfield  {author} {\bibinfo {author} {\bibfnamefont {W.}~\bibnamefont
  {Hu}}, \bibinfo {author} {\bibfnamefont {R.~R.~P.}\ \bibnamefont {Singh}},\
  and\ \bibinfo {author} {\bibfnamefont {R.~T.}\ \bibnamefont {Scalettar}},\
  }\bibfield  {title} {\bibinfo {title} {Discovering phases, phase transitions,
  and crossovers through unsupervised machine learning: A critical
  examination},\ }\href {https://doi.org/10.1103/PhysRevE.95.062122} {\bibfield
   {journal} {\bibinfo  {journal} {Phys. Rev. E}\ }\textbf {\bibinfo {volume}
  {95}},\ \bibinfo {pages} {062122} (\bibinfo {year} {2017})}\BibitemShut
  {NoStop}%
\bibitem [{\citenamefont {Zhang}\ \emph {et~al.}(2019)\citenamefont {Zhang},
  \citenamefont {Liu},\ and\ \citenamefont {Wei}}]{Zhang_W}%
  \BibitemOpen
  \bibfield  {author} {\bibinfo {author} {\bibfnamefont {W.}~\bibnamefont
  {Zhang}}, \bibinfo {author} {\bibfnamefont {J.}~\bibnamefont {Liu}},\ and\
  \bibinfo {author} {\bibfnamefont {T.-C.}\ \bibnamefont {Wei}},\ }\bibfield
  {title} {\bibinfo {title} {Machine learning of phase transitions in the
  percolation and {XY} models},\ }\href
  {https://doi.org/10.1103/PhysRevE.99.032142} {\bibfield  {journal} {\bibinfo
  {journal} {Phys. Rev. E}\ }\textbf {\bibinfo {volume} {99}},\ \bibinfo
  {pages} {032142} (\bibinfo {year} {2019})}\BibitemShut {NoStop}%
\bibitem [{\citenamefont {Ch'ng}\ \emph {et~al.}(2018)\citenamefont {Ch'ng},
  \citenamefont {Vazquez},\ and\ \citenamefont {Khatami}}]{tsne4}%
  \BibitemOpen
  \bibfield  {author} {\bibinfo {author} {\bibfnamefont {K.}~\bibnamefont
  {Ch'ng}}, \bibinfo {author} {\bibfnamefont {N.}~\bibnamefont {Vazquez}},\
  and\ \bibinfo {author} {\bibfnamefont {E.}~\bibnamefont {Khatami}},\
  }\bibfield  {title} {\bibinfo {title} {Unsupervised machine learning account
  of magnetic transitions in the {Hubbard} model},\ }\href
  {https://doi.org/10.1103/PhysRevE.97.013306} {\bibfield  {journal} {\bibinfo
  {journal} {Phys. Rev. E}\ }\textbf {\bibinfo {volume} {97}},\ \bibinfo
  {pages} {013306} (\bibinfo {year} {2018})}\BibitemShut {NoStop}%
\bibitem [{\citenamefont {Yang}\ \emph {et~al.}(2021)\citenamefont {Yang},
  \citenamefont {Sun}, \citenamefont {Ran},\ and\ \citenamefont {Su}}]{tsne3}%
  \BibitemOpen
  \bibfield  {author} {\bibinfo {author} {\bibfnamefont {Y.}~\bibnamefont
  {Yang}}, \bibinfo {author} {\bibfnamefont {Z.-Z.}\ \bibnamefont {Sun}},
  \bibinfo {author} {\bibfnamefont {S.-J.}\ \bibnamefont {Ran}},\ and\ \bibinfo
  {author} {\bibfnamefont {G.}~\bibnamefont {Su}},\ }\bibfield  {title}
  {\bibinfo {title} {Visualizing quantum phases and identifying quantum phase
  transitions by nonlinear dimensional reduction},\ }\href
  {https://doi.org/10.1103/PhysRevB.103.075106} {\bibfield  {journal} {\bibinfo
   {journal} {Phys. Rev. B}\ }\textbf {\bibinfo {volume} {103}},\ \bibinfo
  {pages} {075106} (\bibinfo {year} {2021})}\BibitemShut {NoStop}%
\bibitem [{\citenamefont {Rodriguez-Nieva}\ and\ \citenamefont
  {Scheurer}(2019)}]{df1}%
  \BibitemOpen
  \bibfield  {author} {\bibinfo {author} {\bibfnamefont {J.~F.}\ \bibnamefont
  {Rodriguez-Nieva}}\ and\ \bibinfo {author} {\bibfnamefont {M.~S.}\
  \bibnamefont {Scheurer}},\ }\bibfield  {title} {\bibinfo {title} {Identifying
  topological order through unsupervised machine learning},\ }\href
  {https://www.nature.com/articles/s41567-019-0512-x} {\bibfield  {journal}
  {\bibinfo  {journal} {Nat. Phys.}\ }\textbf {\bibinfo {volume} {15}},\
  \bibinfo {pages} {790} (\bibinfo {year} {2019})}\BibitemShut {NoStop}%
\bibitem [{\citenamefont {Scheurer}\ and\ \citenamefont {Slager}(2020)}]{df2}%
  \BibitemOpen
  \bibfield  {author} {\bibinfo {author} {\bibfnamefont {M.~S.}\ \bibnamefont
  {Scheurer}}\ and\ \bibinfo {author} {\bibfnamefont {R.-J.}\ \bibnamefont
  {Slager}},\ }\bibfield  {title} {\bibinfo {title} {Unsupervised machine
  learning and band topology},\ }\href
  {https://doi.org/10.1103/PhysRevLett.124.226401} {\bibfield  {journal}
  {\bibinfo  {journal} {Phys. Rev. Lett.}\ }\textbf {\bibinfo {volume} {124}},\
  \bibinfo {pages} {226401} (\bibinfo {year} {2020})}\BibitemShut {NoStop}%
\bibitem [{\citenamefont {Lidiak}\ and\ \citenamefont {Gong}(2020)}]{df3}%
  \BibitemOpen
  \bibfield  {author} {\bibinfo {author} {\bibfnamefont {A.}~\bibnamefont
  {Lidiak}}\ and\ \bibinfo {author} {\bibfnamefont {Z.}~\bibnamefont {Gong}},\
  }\bibfield  {title} {\bibinfo {title} {Unsupervised machine learning of
  quantum phase transitions using diffusion maps},\ }\href
  {https://doi.org/10.1103/PhysRevLett.125.225701} {\bibfield  {journal}
  {\bibinfo  {journal} {Phys. Rev. Lett.}\ }\textbf {\bibinfo {volume} {125}},\
  \bibinfo {pages} {225701} (\bibinfo {year} {2020})}\BibitemShut {NoStop}%
\bibitem [{\citenamefont {Long}\ \emph {et~al.}(2020)\citenamefont {Long},
  \citenamefont {Ren},\ and\ \citenamefont {Chen}}]{df4}%
  \BibitemOpen
  \bibfield  {author} {\bibinfo {author} {\bibfnamefont {Y.}~\bibnamefont
  {Long}}, \bibinfo {author} {\bibfnamefont {J.}~\bibnamefont {Ren}},\ and\
  \bibinfo {author} {\bibfnamefont {H.}~\bibnamefont {Chen}},\ }\bibfield
  {title} {\bibinfo {title} {Unsupervised manifold clustering of topological
  phononics},\ }\href {https://doi.org/10.1103/PhysRevLett.124.185501}
  {\bibfield  {journal} {\bibinfo  {journal} {Phys. Rev. Lett.}\ }\textbf
  {\bibinfo {volume} {124}},\ \bibinfo {pages} {185501} (\bibinfo {year}
  {2020})}\BibitemShut {NoStop}%
\bibitem [{\citenamefont {{Van Nieuwenburg}}\ \emph {et~al.}(2017)\citenamefont
  {{Van Nieuwenburg}}, \citenamefont {Liu},\ and\ \citenamefont
  {Huber}}]{confusion}%
  \BibitemOpen
  \bibfield  {author} {\bibinfo {author} {\bibfnamefont {E.~P.~L.}\
  \bibnamefont {{Van Nieuwenburg}}}, \bibinfo {author} {\bibfnamefont {Y.-H.}\
  \bibnamefont {Liu}},\ and\ \bibinfo {author} {\bibfnamefont {S.~D.}\
  \bibnamefont {Huber}},\ }\bibfield  {title} {\bibinfo {title} {Learning phase
  transitions by confusion},\ }\href
  {https://www.nature.com/articles/nphys4037} {\bibfield  {journal} {\bibinfo
  {journal} {Nat. Phys.}\ }\textbf {\bibinfo {volume} {13}},\ \bibinfo {pages}
  {435} (\bibinfo {year} {2017})}\BibitemShut {NoStop}%
\bibitem [{\citenamefont {Cheng}\ \emph {et~al.}(2021)\citenamefont {Cheng},
  \citenamefont {He}, \citenamefont {Zhang}, \citenamefont {Zhu},\ and\
  \citenamefont {Shi}}]{huizhai}%
  \BibitemOpen
  \bibfield  {author} {\bibinfo {author} {\bibfnamefont {S.}~\bibnamefont
  {Cheng}}, \bibinfo {author} {\bibfnamefont {F.}~\bibnamefont {He}}, \bibinfo
  {author} {\bibfnamefont {H.}~\bibnamefont {Zhang}}, \bibinfo {author}
  {\bibfnamefont {K.-D.}\ \bibnamefont {Zhu}},\ and\ \bibinfo {author}
  {\bibfnamefont {Y.}~\bibnamefont {Shi}},\ }\bibfield  {title} {\bibinfo
  {title} {Machine learning percolation model},\ }\href
  {https://arxiv.org/abs/2101.08928} {\bibfield  {journal} {\bibinfo  {journal}
  {arXiv preprint arXiv:2101.08928}\ } (\bibinfo {year} {2021})}\BibitemShut
  {NoStop}%
\bibitem [{\citenamefont {Xu}\ \emph {et~al.}(2019)\citenamefont {Xu},
  \citenamefont {Fu},\ and\ \citenamefont {Zhao}}]{hzhao}%
  \BibitemOpen
  \bibfield  {author} {\bibinfo {author} {\bibfnamefont {R.}~\bibnamefont
  {Xu}}, \bibinfo {author} {\bibfnamefont {W.}~\bibnamefont {Fu}},\ and\
  \bibinfo {author} {\bibfnamefont {H.}~\bibnamefont {Zhao}},\ }\bibfield
  {title} {\bibinfo {title} {A new strategy in applying the learning machine to
  study phase transitions},\ }\href {https://arxiv.org/abs/1901.00774}
  {\bibfield  {journal} {\bibinfo  {journal} {arXiv preprint arXiv:1901.00774}\
  } (\bibinfo {year} {2019})}\BibitemShut {NoStop}%
\bibitem [{\citenamefont {Wang}\ \emph {et~al.}(2021)\citenamefont {Wang},
  \citenamefont {Zhang}, \citenamefont {Hua},\ and\ \citenamefont
  {Wei}}]{PhysRevResearch.3.013074}%
  \BibitemOpen
  \bibfield  {author} {\bibinfo {author} {\bibfnamefont {J.}~\bibnamefont
  {Wang}}, \bibinfo {author} {\bibfnamefont {W.}~\bibnamefont {Zhang}},
  \bibinfo {author} {\bibfnamefont {T.}~\bibnamefont {Hua}},\ and\ \bibinfo
  {author} {\bibfnamefont {T.-C.}\ \bibnamefont {Wei}},\ }\bibfield  {title}
  {\bibinfo {title} {Unsupervised learning of topological phase transitions
  using the {Calinski-Harabaz} index},\ }\href
  {https://doi.org/10.1103/PhysRevResearch.3.013074} {\bibfield  {journal}
  {\bibinfo  {journal} {Phys. Rev. Research}\ }\textbf {\bibinfo {volume}
  {3}},\ \bibinfo {pages} {013074} (\bibinfo {year} {2021})}\BibitemShut
  {NoStop}%
\bibitem [{\citenamefont {Suchsland}\ and\ \citenamefont
  {Wessel}(2018)}]{wessel}%
  \BibitemOpen
  \bibfield  {author} {\bibinfo {author} {\bibfnamefont {P.}~\bibnamefont
  {Suchsland}}\ and\ \bibinfo {author} {\bibfnamefont {S.}~\bibnamefont
  {Wessel}},\ }\bibfield  {title} {\bibinfo {title} {Parameter diagnostics of
  phases and phase transition learning by neural networks},\ }\href
  {https://doi.org/10.1103/PhysRevB.97.174435} {\bibfield  {journal} {\bibinfo
  {journal} {Phys. Rev. B}\ }\textbf {\bibinfo {volume} {97}},\ \bibinfo
  {pages} {174435} (\bibinfo {year} {2018})}\BibitemShut {NoStop}%
\bibitem [{\citenamefont {Lee}\ and\ \citenamefont {Kim}(2019)}]{cf_qlr}%
  \BibitemOpen
  \bibfield  {author} {\bibinfo {author} {\bibfnamefont {S.~S.}\ \bibnamefont
  {Lee}}\ and\ \bibinfo {author} {\bibfnamefont {B.~J.}\ \bibnamefont {Kim}},\
  }\bibfield  {title} {\bibinfo {title} {{Confusion scheme in machine learning
  detects double phase transitions and quasi-long-range order}},\ }\href
  {https://doi.org/10.1103/PhysRevE.99.043308} {\bibfield  {journal} {\bibinfo
  {journal} {Phys. Rev. E}\ }\textbf {\bibinfo {volume} {99}},\ \bibinfo
  {pages} {43308} (\bibinfo {year} {2019})}\BibitemShut {NoStop}%
\bibitem [{\citenamefont {Jacobsen}(2015)}]{Jacobsen_2015}%
  \BibitemOpen
  \bibfield  {author} {\bibinfo {author} {\bibfnamefont {J.~L.}\ \bibnamefont
  {Jacobsen}},\ }\bibfield  {title} {\bibinfo {title} {Critical points of potts
  and {O(N)} models from eigenvalue identities in periodic
  {Temperley{\textendash}Lieb} algebras},\ }\href
  {https://doi.org/10.1088/1751-8113/48/45/454003} {\bibfield  {journal}
  {\bibinfo  {journal} {J. Phys. A Math. Theor.}\ }\textbf {\bibinfo {volume}
  {48}},\ \bibinfo {pages} {454003} (\bibinfo {year} {2015})}\BibitemShut
  {NoStop}%
\bibitem [{\citenamefont {Feng}\ \emph {et~al.}(2008)\citenamefont {Feng},
  \citenamefont {Deng},\ and\ \citenamefont {Bl\"ote}}]{PhysRevE.78.031136}%
  \BibitemOpen
  \bibfield  {author} {\bibinfo {author} {\bibfnamefont {X.}~\bibnamefont
  {Feng}}, \bibinfo {author} {\bibfnamefont {Y.}~\bibnamefont {Deng}},\ and\
  \bibinfo {author} {\bibfnamefont {H.~W.~J.}\ \bibnamefont {Bl\"ote}},\
  }\bibfield  {title} {\bibinfo {title} {Percolation transitions in two
  dimensions},\ }\href {https://doi.org/10.1103/PhysRevE.78.031136} {\bibfield
  {journal} {\bibinfo  {journal} {Phys. Rev. E}\ }\textbf {\bibinfo {volume}
  {78}},\ \bibinfo {pages} {031136} (\bibinfo {year} {2008})}\BibitemShut
  {NoStop}%
\bibitem [{\citenamefont {Sykes}\ and\ \citenamefont
  {Essam}(1964)}]{doi:10.1063/1.1704215}%
  \BibitemOpen
  \bibfield  {author} {\bibinfo {author} {\bibfnamefont {M.~F.}\ \bibnamefont
  {Sykes}}\ and\ \bibinfo {author} {\bibfnamefont {J.~W.}\ \bibnamefont
  {Essam}},\ }\bibfield  {title} {\bibinfo {title} {Exact critical percolation
  probabilities for site and bond problems in two dimensions},\ }\href
  {https://doi.org/10.1063/1.1704215} {\bibfield  {journal} {\bibinfo
  {journal} {J. Math. Phys.}\ }\textbf {\bibinfo {volume} {5}},\ \bibinfo
  {pages} {1117} (\bibinfo {year} {1964})}\BibitemShut {NoStop}%
\bibitem [{\citenamefont {Jacobsen}(2014)}]{threshold2}%
  \BibitemOpen
  \bibfield  {author} {\bibinfo {author} {\bibfnamefont {J.~L.}\ \bibnamefont
  {Jacobsen}},\ }\bibfield  {title} {\bibinfo {title} {High-precision
  percolation thresholds and {Potts}-model critical manifolds from graph
  polynomials},\ }\href {https://doi.org/10.1088/1751-8113/47/13/135001}
  {\bibfield  {journal} {\bibinfo  {journal} {J. Phys. A Math. Theor.}\
  }\textbf {\bibinfo {volume} {47}},\ \bibinfo {pages} {135001} (\bibinfo
  {year} {2014})}\BibitemShut {NoStop}%
\bibitem [{\citenamefont {Kesten}(1980)}]{kesten1980critical}%
  \BibitemOpen
  \bibfield  {author} {\bibinfo {author} {\bibfnamefont {H.}~\bibnamefont
  {Kesten}},\ }\bibfield  {title} {\bibinfo {title} {The critical probability
  of bond percolation on the square lattice equals 1/2},\ }\href
  {https://doi.org/10.1007/BF01197577} {\bibfield  {journal} {\bibinfo
  {journal} {Commun. Math. Phys.}\ }\textbf {\bibinfo {volume} {74}},\ \bibinfo
  {pages} {41} (\bibinfo {year} {1980})}\BibitemShut {NoStop}%
\bibitem [{\citenamefont {Erdos}\ and\ \citenamefont
  {R{\'e}nyi}(1960)}]{erdos1960evolution}%
  \BibitemOpen
  \bibfield  {author} {\bibinfo {author} {\bibfnamefont {P.}~\bibnamefont
  {Erdos}}\ and\ \bibinfo {author} {\bibfnamefont {A.}~\bibnamefont
  {R{\'e}nyi}},\ }\bibfield  {title} {\bibinfo {title} {On the evolution of
  random graphs},\ }\href@noop {} {\bibfield  {journal} {\bibinfo  {journal}
  {Publ. Math. Inst. Hung. Acad. Sci}\ }\textbf {\bibinfo {volume} {5}},\
  \bibinfo {pages} {17} (\bibinfo {year} {1960})}\BibitemShut {NoStop}%
\bibitem [{\citenamefont {Malarz}\ and\ \citenamefont
  {Galam}(2005)}]{8neighbor}%
  \BibitemOpen
  \bibfield  {author} {\bibinfo {author} {\bibfnamefont {K.}~\bibnamefont
  {Malarz}}\ and\ \bibinfo {author} {\bibfnamefont {S.}~\bibnamefont {Galam}},\
  }\bibfield  {title} {\bibinfo {title} {Square-lattice site percolation at
  increasing ranges of neighbor bonds},\ }\href
  {https://doi.org/10.1103/PhysRevE.71.016125} {\bibfield  {journal} {\bibinfo
  {journal} {Phys. Rev. E}\ }\textbf {\bibinfo {volume} {71}},\ \bibinfo
  {pages} {16125} (\bibinfo {year} {2005})}\BibitemShut {NoStop}%
\bibitem [{\citenamefont {Majewski}\ and\ \citenamefont
  {Malarz}(2007)}]{ActaPhysPB}%
  \BibitemOpen
  \bibfield  {author} {\bibinfo {author} {\bibfnamefont {M.}~\bibnamefont
  {Majewski}}\ and\ \bibinfo {author} {\bibfnamefont {K.}~\bibnamefont
  {Malarz}},\ }\bibfield  {title} {\bibinfo {title} {Square lattice site
  percolation thresholds for complex neighbourhoods},\ }\href
  {https://www.actaphys.uj.edu.pl/R/38/6/2191/pdf} {\bibfield  {journal}
  {\bibinfo  {journal} {Acta Phys. Pol. B}\ }\textbf {\bibinfo {volume} {38}},\
  \bibinfo {pages} {2191} (\bibinfo {year} {2007})}\BibitemShut {NoStop}%
\bibitem [{\citenamefont {Ferrenberg}\ and\ \citenamefont
  {Landau}(1991)}]{landau}%
  \BibitemOpen
  \bibfield  {author} {\bibinfo {author} {\bibfnamefont {A.~M.}\ \bibnamefont
  {Ferrenberg}}\ and\ \bibinfo {author} {\bibfnamefont {D.~P.}\ \bibnamefont
  {Landau}},\ }\bibfield  {title} {\bibinfo {title} {Critical behavior of the
  three-dimensional ising model: A high-resolution monte carlo study},\ }\href
  {https://doi.org/10.1103/PhysRevB.44.5081} {\bibfield  {journal} {\bibinfo
  {journal} {Phys. Rev. B}\ }\textbf {\bibinfo {volume} {44}},\ \bibinfo
  {pages} {5081} (\bibinfo {year} {1991})}\BibitemShut {NoStop}%
\bibitem [{\citenamefont {Tan}\ \emph {et~al.}(2020)\citenamefont {Tan},
  \citenamefont {Li}, \citenamefont {Zhu},\ and\ \citenamefont
  {Jiang}}]{tan2020comprehensive}%
  \BibitemOpen
  \bibfield  {author} {\bibinfo {author} {\bibfnamefont {D.-R.}\ \bibnamefont
  {Tan}}, \bibinfo {author} {\bibfnamefont {C.-D.}\ \bibnamefont {Li}},
  \bibinfo {author} {\bibfnamefont {W.-P.}\ \bibnamefont {Zhu}},\ and\ \bibinfo
  {author} {\bibfnamefont {F.-J.}\ \bibnamefont {Jiang}},\ }\bibfield  {title}
  {\bibinfo {title} {A comprehensive neural networks study of the phase
  transitions of {Potts} model},\ }\href
  {https://doi.org/10.1088/1367-2630/ab8ab4} {\bibfield  {journal} {\bibinfo
  {journal} {New J. Phys.}\ }\textbf {\bibinfo {volume} {22}},\ \bibinfo
  {pages} {063016} (\bibinfo {year} {2020})}\BibitemShut {NoStop}%
\bibitem [{\citenamefont {Li}\ \emph {et~al.}(2018)\citenamefont {Li},
  \citenamefont {Tan},\ and\ \citenamefont {Jiang}}]{li2018applications}%
  \BibitemOpen
  \bibfield  {author} {\bibinfo {author} {\bibfnamefont {C.-D.}\ \bibnamefont
  {Li}}, \bibinfo {author} {\bibfnamefont {D.-R.}\ \bibnamefont {Tan}},\ and\
  \bibinfo {author} {\bibfnamefont {F.-J.}\ \bibnamefont {Jiang}},\ }\bibfield
  {title} {\bibinfo {title} {Applications of neural networks to the studies of
  phase transitions of two-dimensional {Potts} models},\ }\href
  {https://doi.org/https://doi.org/10.1016/j.aop.2018.02.018} {\bibfield
  {journal} {\bibinfo  {journal} {Ann. Phys.}\ }\textbf {\bibinfo {volume}
  {391}},\ \bibinfo {pages} {312} (\bibinfo {year} {2018})}\BibitemShut
  {NoStop}%
\bibitem [{\citenamefont {Shiina}\ \emph {et~al.}(2020)\citenamefont {Shiina},
  \citenamefont {Mori}, \citenamefont {Okabe},\ and\ \citenamefont
  {Lee}}]{potts}%
  \BibitemOpen
  \bibfield  {author} {\bibinfo {author} {\bibfnamefont {K.}~\bibnamefont
  {Shiina}}, \bibinfo {author} {\bibfnamefont {H.}~\bibnamefont {Mori}},
  \bibinfo {author} {\bibfnamefont {Y.}~\bibnamefont {Okabe}},\ and\ \bibinfo
  {author} {\bibfnamefont {H.~K.}\ \bibnamefont {Lee}},\ }\bibfield  {title}
  {\bibinfo {title} {{Machine-learning studies on spin models}},\ }\href
  {https://www.nature.com/articles/s41598-020-58263-5} {\bibfield  {journal}
  {\bibinfo  {journal} {Sci. Rep.}\ }\textbf {\bibinfo {volume} {10}},\
  \bibinfo {pages} {1} (\bibinfo {year} {2020})}\BibitemShut {NoStop}%
\bibitem [{\citenamefont {Hu}\ \emph {et~al.}(2011)\citenamefont {Hu},
  \citenamefont {Deng},\ and\ \citenamefont {Bl\"ote}}]{perco-xy1}%
  \BibitemOpen
  \bibfield  {author} {\bibinfo {author} {\bibfnamefont {H.}~\bibnamefont
  {Hu}}, \bibinfo {author} {\bibfnamefont {Y.}~\bibnamefont {Deng}},\ and\
  \bibinfo {author} {\bibfnamefont {H.~W.~J.}\ \bibnamefont {Bl\"ote}},\
  }\bibfield  {title} {\bibinfo {title} {{Berezinskii-Kosterlitz-Thouless}-like
  percolation transitions in the two-dimensional {XY} model},\ }\href
  {https://doi.org/10.1103/PhysRevE.83.011124} {\bibfield  {journal} {\bibinfo
  {journal} {Phys. Rev. E}\ }\textbf {\bibinfo {volume} {83}},\ \bibinfo
  {pages} {011124} (\bibinfo {year} {2011})}\BibitemShut {NoStop}%
\bibitem [{\citenamefont {Hintermann}\ \emph {et~al.}(1978)\citenamefont
  {Hintermann}, \citenamefont {Kunz},\ and\ \citenamefont
  {Wu}}]{Hintermann1978}%
  \BibitemOpen
  \bibfield  {author} {\bibinfo {author} {\bibfnamefont {A.}~\bibnamefont
  {Hintermann}}, \bibinfo {author} {\bibfnamefont {H.}~\bibnamefont {Kunz}},\
  and\ \bibinfo {author} {\bibfnamefont {F.~Y.}\ \bibnamefont {Wu}},\
  }\bibfield  {title} {\bibinfo {title} {Exact results for the {Potts} model in
  two dimensions},\ }\href {https://doi.org/10.1007/BF01011773} {\bibfield
  {journal} {\bibinfo  {journal} {J. Stat. Phys.}\ }\textbf {\bibinfo {volume}
  {19}},\ \bibinfo {pages} {623} (\bibinfo {year} {1978})}\BibitemShut
  {NoStop}%
\bibitem [{\citenamefont {Kosterlitz}\ and\ \citenamefont
  {Thouless}(1973)}]{BKT}%
  \BibitemOpen
  \bibfield  {author} {\bibinfo {author} {\bibfnamefont {J.~M.}\ \bibnamefont
  {Kosterlitz}}\ and\ \bibinfo {author} {\bibfnamefont {D.~J.}\ \bibnamefont
  {Thouless}},\ }\bibfield  {title} {\bibinfo {title} {Ordering, metastability
  and phase transitions in two-dimensional systems},\ }\href
  {https://doi.org/10.1088/0022-3719/6/7/010} {\bibfield  {journal} {\bibinfo
  {journal} {J. Phys. C}\ }\textbf {\bibinfo {volume} {6}},\ \bibinfo {pages}
  {1181} (\bibinfo {year} {1973})}\BibitemShut {NoStop}%
\bibitem [{\citenamefont {Hsieh}\ \emph {et~al.}(2013)\citenamefont {Hsieh},
  \citenamefont {Kao},\ and\ \citenamefont {Sandvik}}]{Hsieh_2013}%
  \BibitemOpen
  \bibfield  {author} {\bibinfo {author} {\bibfnamefont {Y.-D.}\ \bibnamefont
  {Hsieh}}, \bibinfo {author} {\bibfnamefont {Y.-J.}\ \bibnamefont {Kao}},\
  and\ \bibinfo {author} {\bibfnamefont {A.~W.}\ \bibnamefont {Sandvik}},\
  }\bibfield  {title} {\bibinfo {title} {Finite-size scaling method for the
  {Berezinskii{\textendash}Kosterlitz{\textendash}Thouless} transition},\
  }\href {https://doi.org/10.1088/1742-5468/2013/09/p09001} {\bibfield
  {journal} {\bibinfo  {journal} {J. Stat. Mech.}\ }\textbf {\bibinfo {volume}
  {2013}},\ \bibinfo {pages} {P09001} (\bibinfo {year} {2013})}\BibitemShut
  {NoStop}%
\bibitem [{\citenamefont {Kosterlitz}(1974)}]{Kosterlitz_1974}%
  \BibitemOpen
  \bibfield  {author} {\bibinfo {author} {\bibfnamefont {J.~M.}\ \bibnamefont
  {Kosterlitz}},\ }\bibfield  {title} {\bibinfo {title} {The critical
  properties of the two-dimensional xy model},\ }\href
  {https://doi.org/10.1088/0022-3719/7/6/005} {\bibfield  {journal} {\bibinfo
  {journal} {J. Phys. C}\ }\textbf {\bibinfo {volume} {7}},\ \bibinfo {pages}
  {1046} (\bibinfo {year} {1974})}\BibitemShut {NoStop}%
\bibitem [{\citenamefont {Weber}\ and\ \citenamefont
  {Minnhagen}(1988)}]{PhysRevB.37.5986}%
  \BibitemOpen
  \bibfield  {author} {\bibinfo {author} {\bibfnamefont {H.}~\bibnamefont
  {Weber}}\ and\ \bibinfo {author} {\bibfnamefont {P.}~\bibnamefont
  {Minnhagen}},\ }\bibfield  {title} {\bibinfo {title} {Monte carlo
  determination of the critical temperature for the two-dimensional xy model},\
  }\href {https://doi.org/10.1103/PhysRevB.37.5986} {\bibfield  {journal}
  {\bibinfo  {journal} {Phys. Rev. B}\ }\textbf {\bibinfo {volume} {37}},\
  \bibinfo {pages} {5986} (\bibinfo {year} {1988})}\BibitemShut {NoStop}%
\bibitem [{\citenamefont {Fisher}(1998)}]{universality}%
  \BibitemOpen
  \bibfield  {author} {\bibinfo {author} {\bibfnamefont {M.~E.}\ \bibnamefont
  {Fisher}},\ }\bibfield  {title} {\bibinfo {title} {Renormalization group
  theory: Its basis and formulation in statistical physics},\ }\href
  {https://doi.org/10.1103/RevModPhys.70.653} {\bibfield  {journal} {\bibinfo
  {journal} {Rev. Mod. Phys}\ }\textbf {\bibinfo {volume} {70}},\ \bibinfo
  {pages} {653} (\bibinfo {year} {1998})}\BibitemShut {NoStop}%
\bibitem [{\citenamefont {Li}\ \emph {et~al.}(2019)\citenamefont {Li},
  \citenamefont {Luo},\ and\ \citenamefont {Wan}}]{Critical_Exponent1}%
  \BibitemOpen
  \bibfield  {author} {\bibinfo {author} {\bibfnamefont {Z.}~\bibnamefont
  {Li}}, \bibinfo {author} {\bibfnamefont {M.}~\bibnamefont {Luo}},\ and\
  \bibinfo {author} {\bibfnamefont {X.}~\bibnamefont {Wan}},\ }\bibfield
  {title} {\bibinfo {title} {Extracting critical exponents by finite-size
  scaling with convolutional neural networks},\ }\href
  {https://doi.org/10.1103/PhysRevB.99.075418} {\bibfield  {journal} {\bibinfo
  {journal} {Phys. Rev. B}\ }\textbf {\bibinfo {volume} {99}},\ \bibinfo
  {pages} {75418} (\bibinfo {year} {2019})}\BibitemShut {NoStop}%
\bibitem [{\citenamefont {Giannetti}\ \emph {et~al.}(2019)\citenamefont
  {Giannetti}, \citenamefont {Lucini},\ and\ \citenamefont
  {Vadacchino}}]{Giannetti}%
  \BibitemOpen
  \bibfield  {author} {\bibinfo {author} {\bibfnamefont {C.}~\bibnamefont
  {Giannetti}}, \bibinfo {author} {\bibfnamefont {B.}~\bibnamefont {Lucini}},\
  and\ \bibinfo {author} {\bibfnamefont {D.}~\bibnamefont {Vadacchino}},\
  }\bibfield  {title} {\bibinfo {title} {Machine learning as a universal tool
  for quantitative investigations of phase transitions},\ }\href
  {https://doi.org/10.1016/j.nuclphysb.2019.114639} {\bibfield  {journal}
  {\bibinfo  {journal} {Nucl. Phys. B.}\ }\textbf {\bibinfo {volume} {944}},\
  \bibinfo {pages} {114639} (\bibinfo {year} {2019})}\BibitemShut {NoStop}%
\bibitem [{\citenamefont {Baxter}(1973)}]{baxter1973potts}%
  \BibitemOpen
  \bibfield  {author} {\bibinfo {author} {\bibfnamefont {R.~J.}\ \bibnamefont
  {Baxter}},\ }\bibfield  {title} {\bibinfo {title} {{Potts} model at the
  critical temperature},\ }\href
  {https://iopscience.iop.org/article/10.1088/0022-3719/6/23/005} {\bibfield
  {journal} {\bibinfo  {journal} {J. Phys. C}\ }\textbf {\bibinfo {volume}
  {6}},\ \bibinfo {pages} {L445} (\bibinfo {year} {1973})}\BibitemShut
  {NoStop}%
\bibitem [{\citenamefont {Baxter}\ \emph {et~al.}(1978)\citenamefont {Baxter},
  \citenamefont {Temperley}, \citenamefont {Ashley},\ and\ \citenamefont
  {Edwards}}]{baxter1978triangular}%
  \BibitemOpen
  \bibfield  {author} {\bibinfo {author} {\bibfnamefont {R.~J.}\ \bibnamefont
  {Baxter}}, \bibinfo {author} {\bibfnamefont {H.~N.~V.}\ \bibnamefont
  {Temperley}}, \bibinfo {author} {\bibfnamefont {S.~E.}\ \bibnamefont
  {Ashley}},\ and\ \bibinfo {author} {\bibfnamefont {S.}~\bibnamefont
  {Edwards}},\ }\bibfield  {title} {\bibinfo {title} {Triangular {Potts} model
  at its transition temperature, and related models},\ }\href
  {https://doi.org/10.1098/rspa.1978.0026} {\bibfield  {journal} {\bibinfo
  {journal} {Proc. Math. Phys. Sci.}\ }\textbf {\bibinfo {volume} {358}},\
  \bibinfo {pages} {535} (\bibinfo {year} {1978})}\BibitemShut {NoStop}%
\bibitem [{\citenamefont {Goodfellow}\ \emph {et~al.}(2016)\citenamefont
  {Goodfellow}, \citenamefont {Bengio},\ and\ \citenamefont
  {Courville}}]{goodfellow2016deep}%
  \BibitemOpen
  \bibfield  {author} {\bibinfo {author} {\bibfnamefont {I.}~\bibnamefont
  {Goodfellow}}, \bibinfo {author} {\bibfnamefont {Y.}~\bibnamefont {Bengio}},\
  and\ \bibinfo {author} {\bibfnamefont {A.}~\bibnamefont {Courville}},\ }\href
  {https://doi.org/10.1007/s10710-017-9314-z} {\emph {\bibinfo {title} {Deep
  learning}}}\ (\bibinfo  {publisher} {MIT press},\ \bibinfo {year}
  {2016})\BibitemShut {NoStop}%
\end{thebibliography}%
\end{document}